\documentclass{article}

\usepackage{PRIMEarxiv}

\usepackage{subfigure}
\usepackage{multirow}
\usepackage{threeparttable}
\usepackage{tabularx} 

\usepackage[utf8]{inputenc} % allow utf-8 input
\usepackage[T1]{fontenc}    % use 8-bit T1 fonts
\usepackage{hyperref}       % hyperlinks
\usepackage{url}            % simple URL typesetting
\usepackage{booktabs}       % professional-quality tables
\usepackage{amsfonts}       % blackboard math symbols
\usepackage{nicefrac}       % compact symbols for 1/2, etc.
\usepackage{microtype}      % microtypography
\usepackage{lipsum}
\usepackage{fancyhdr}       % header
\usepackage{graphicx}       % graphics
\graphicspath{{media/}}     % organize your images and other figures under media/ folder

\title{Selective Complementary Feature Fusion and Modal Feature Compression Interaction for Brain Tumor Segmentation
}

\author{
  Dong Chen \quad Boyue Zhao \quad Yi Zhang \quad Meng Zhao\thanks{Corresponding author: Meng Zhao} \\
  Engineering Research Center of Learning-Based Intelligent System \thanks{The Engineering Research Center of Learning-Based Intelligent System is in the Key Laboratory of Computer Vision and System of Ministry of Education, certified by the Ministry of Education} \\
  Tianjin University of Technology \\
  Tianjin City\\
  \texttt{\{Dong Chen, Boyue Zhao, Yi Zhang, Meng Zhao\}e-mail: cdmm0903@163.com;} \\
  \texttt{zby1004@stud.tjut.edu.cn; wazy1124@stud.tjut.edu.cn; zh\_m@tju.edu.cn}
}

\begin{document}
\maketitle

\begin{abstract}
Efficient modal feature fusion strategy is the key to achieve accurate segmentation of brain glioma. However, due to the specificity of different MRI modes, it is difficult to carry out cross-modal fusion with large differences in modal features, resulting in the model ignoring rich feature information. On the other hand, the problem of multi-modal feature redundancy interaction occurs in parallel networks due to the proliferation of feature dimensions, further increase the difficulty of multi-modal feature fusion at the bottom end.
In order to solve the above problems, we propose a noval complementary feature compression interaction network (CFCI-Net), which realizes the complementary fusion and compression interaction of multi-modal feature information with an efficient mode fusion strategy.
Firstly, we propose a selective complementary feature fusion (SCFF) module, which adaptively fuses rich cross-modal feature information by complementary soft selection weights. 
Secondly, a modal feature compression interaction (MFCI) transformer is proposed to deal with the multi-mode fusion redundancy problem when the feature dimension surges. The MFCI transformer is composed of modal feature compression (MFC) and modal feature interaction (MFI) to realize redundancy feature compression and multi-mode feature interactive learning.
%In MFI, we propose a hierarchical interactive attention mechanism based on multi-head attention.
Evaluations on the BraTS2019 and BraTS2020 datasets demonstrate that CFCI-Net achieves superior results compared to state-of-the-art models.
Code: \href{https://github.com/CDmm0/CFCI-Net}{https://github.com/CDmm0/CFCI-Net}
\end{abstract}

% keywords can be removed

\section{Introduction}
% \label{sec:introduction}
Brain glioma is a common tumor disease in the world, which poses a great threat to human health \cite{menze2014multimodal}. In MRI medical images, especially multimodal MRI brain images, can provide important help in the segmentation of brain glioma. These include T1-weighted(T1), contrast enhanced T1-weighted(T1ce), T2-weighted(T2) and fluid attention inversion recovery(FLAIR) images \cite{ranjbarzadeh2023brain}. Different modal images will provide different degree of characteristic information for different focal areas of the tumor. For example, T1 images help to show the boundary between the tumor and surrounding normal brain tissue, at the same time, T1 and T1ce can show normal brain tissue areas, T2 images help to show the necrotic tissue within the tumor and Flair clearly shows the boundaries of the tumor, which is of great significance for peritumoral edema (ED), non-enhancing tumor(NCR/NET)and enhancing tumor(ET) segmentation tasks \cite{zhou2019review}.

%CNN
In recent years, with the rise of convolutional neural networks, many excellent glioma medical image segmentation models have appeared. For example, 3D U-Net \cite{cciccek20163dunet} with U-shaped structure, encoder, decoder and bridge connector and fully connected Convolutional network (FCN) can complete the 3D segmentation task \cite{kamnitsas2017fcn}, as well as Unet++ \cite{zhou2019unet++}, which appears later. nnUnet \cite{isensee2021nnunet} models are excellent studies that have made important achievements in the field of brain glioma segmentation through convolutional structures. 
%Transformer
Subsequently, with the popularity of Transformer \cite{vaswani2017transformer} model in the field of natural language processing, models using Transformer as an encoder are also applied in the field of computer vision image. For example, ViT \cite{dosovitskiy2020vit} and TransUNet \cite{chen2021transunet} are both excellent applications of Transformer in computer vision. 
% Modal Fusion Strategies
% 互补模态SCFF
With the development of these backbone models, multiple model fusion strategies are gradually applied to brain glioma segmentation, more and more people began to pay attention to the characteristics of the modes to design the network structure, including the brain glioma segmentation models based on dominant edge feature information as a single pathway \cite{zhu2023edge}, based on clinically driven knowledge, feature learning is provided using two sets of modes with similar modal features \cite{lin2023ckd}, or SMG-BTS \cite{han2023smg-bts} uses a modal feature as a spatial guide to achieve accurate segmentation of multi-branch parallel running. However, the feature information extracted from a set of similar or single MRI modes is often not rich enough, which easily makes the model pay too much attention to edge information and single features, and it is difficult to cope with the segmentation task of important lesion areas of ET and TC.
% 压缩交互 MFCI
For other modal feature fusion strategies, stacking multiple modes for model feature learning at the early stage of fusion \cite{wang2021clcu}, and the features of each mode are extracted separately and combined by a separate encoder \cite{havaei2016hemis}, and multiple parallel branches are designed for modal feature fusion \cite{cho2022hybrid}. However, these models do not make full use of the features and connections of each mode for modal feature learning and fusion, especially in parallel networks, it is difficult to deal with the multi-modal interaction fusion under the excitation of the bottom feature dimension. 

% Challenges
In the previous glioma segmentation models, there is much room for improvement in the multimodal feature fusion method. The main challenges come from cross-modal feature extraction with large difference in feature information, and multi-modal learning interaction in parallel network architecture under the condition of channel dimension explosion.
% CFCI-Net
To address these problems, we propose a new brain glioma segmentation network, CFCI-Net, to address the challenges of multimodal feature fusion through cross-modal feature complementary learning and compression interactive fusion.
% SCFF
SCFF cross-modal selective feature fusion method is based on the complementarity of modal features, which uses complementary soft selection weights to make full use of the feature information of each mode. It is worth noting that this complementary soft selection involves pairing modes with significant differences to promote selectivity and complementary feature fusion, thus enriching the modal feature representation of the model and improving the segmentation effect of the model.
% MFCI
To enable efficient multimodal feature interaction learning in the face of feature dimension proliferation, CFCI-Net integrates a modal feature compression interaction (MFCI) transformer that reduces redundant information by comcompression of feature channel dimensions. At the same time, in order to prevent the loss of key feature information, inspired by multi-head attention \cite{vaswani2017transformer}, we integrated an interactive attention mechanism to facilitate multi-modal feature interaction. This iterative feature learning and extraction process, including pre - and post-compression, aims to continuously capture sequence feature information, thereby optimizing the utilization of key modal features and improving the segmentation accuracy of the model.
The proposed CFCI-Net model was rigorously evaluated and validated on the BraTS2019 and BraTS2020 datasets using a comprehensive benchmarking approach. In addition, it has been compared to contemporary open source medical segmentation models, demonstrating its superior performance and establishing its position at the forefront of advanced glioma segmentation methods.

\begin{itemize}
    \item We propose a new segmentation model network, CFCI-Net, to address the challenges of multimodal fusion strategies. It can efficiently fuse cross-modal feature information and cope with multi-modal interaction tasks under the condition of feature proliferation.
    \item We propose SCFF(Selective Complementary Feature Fusion) module, which can adaptively and selectively process the learning weights between the cross-modal features, make full use of the rich information of the cross-modal features, and improve the accuracy of the model segmentation ability.
    \item We propose MFCI(Modal Feature Compression Interaction) transformer to deal with the redundancy problem caused by the parallel structure at the bottom of the model and strengthen the multi-mode fusion interaction. The task of the transformer assembly is to solve the redundancy of feature information and simplify the feature dimensions of the model through the MFC(Modal feature compression) module. In addition, a modal feature interaction (MFI) module and a hierarchical interactive attention mechanism based on multi-head attention mechanism are proposed. MFI uses the interactive attention mechanism to promote the interaction of different modal feature information and improve the robustness and accuracy of the model algorithm performance.
\end{itemize}

\section{Related Works}
\label{sec:headings}
In this section, we will give a brief overview of the important models in the field of brain glioma segmentation, as well as the current work on multimodal segmentation, and we will also introduce the methodological strategies for the fusion of different modes.
\begin{figure*}[t!]
\centering{\includegraphics[width=1\linewidth]{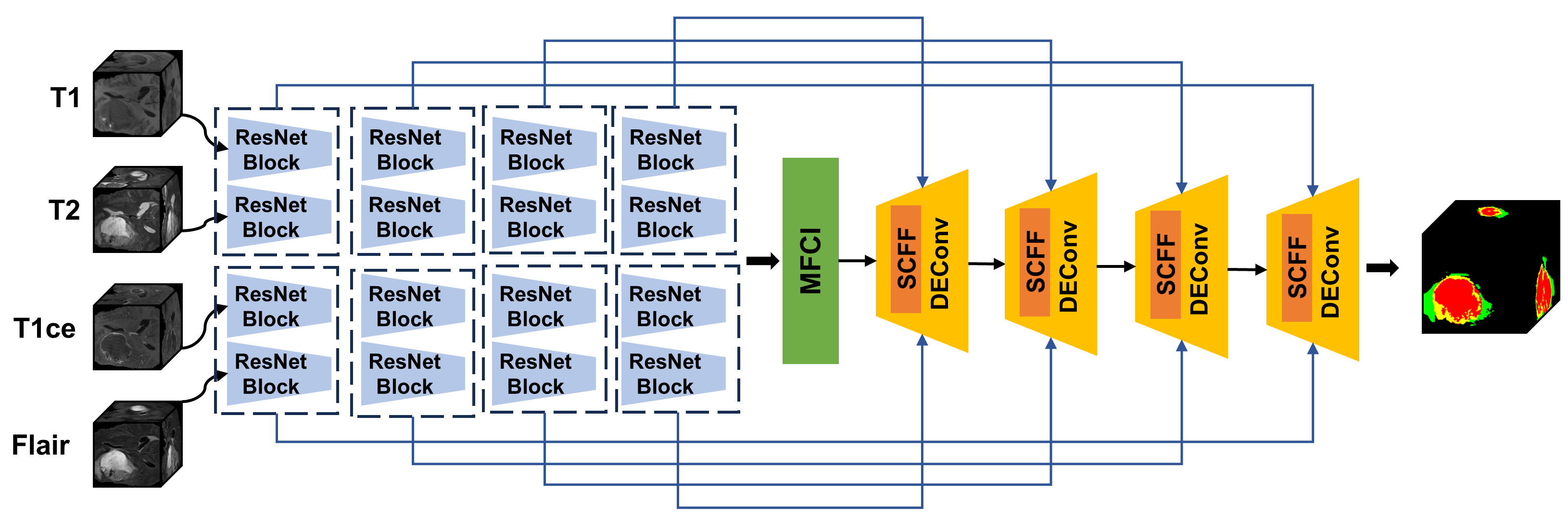}}
\caption{The architecture of the proposed CFCI-Net. The whole model consists of four parallel encoders, MFCI transformer, SCFF module and decoder. The encoder part sets the mode feature extraction path for each mode and is composed of ResNet module. The decoder consists of convolution and upsampling interpolation.}
\label{fig1}
\end{figure*}
\subsection{Previous BTS Models}
With the development of computer vision, there are many outstanding segmentation models for medical image segmentation. From the beginning with the emergence of CNN, U-net \cite{ronneberger20152dunet}, Unet++ \cite{zhou2019unet++} convolutional segmentation networks with 2D images as datasets have achieved remarkable success, and bridge connectors have appeared in many segmentation models to capture long-distance semantic information. Later, nnU-net \cite{isensee2021nnunet}, which uses 3D stereoscopic images as a dataset, was applied in the field of brain glioma segmentation. OM-Net \cite{zhou2020onenet} proposes a new lightweight model that aggregates disparate but related split tasks into a single model. Point-net \cite{ho2021point} combined Point cloud and attention feature map for point volume segmentation according to the characteristics of 3D pixels. CANet \cite{liu2021canet} is a perceptive network that can selectively aggregate contextual feature information. S$^{2}$CA-Net \cite{zhou2024s2ca} enhanced the model's perception of tumor shape and size, harmonizing their common awareness, and thus performing accurate segmentation of brain gliomas.

At the same time, the excellent Transformer model in NLP is also widely used in computer vision. TransBTS \cite{wenxuan2021transbts} has been successful in the field of brain glioma segmentation with global attention. mmFormer \cite{zhang2022mmformer} is a new multimodal medical Transformer used for incomplete multimodal learning for multimodal brain tumor segmentation using Transformer. Swin UNETR \cite{hatamizadeh2021swinunetr} uses sliding window to capture remote feature information for segmentation models. It also retains the U-shaped network structure like U-Net. NestedFormer \cite{xing2022nestedformer} uses PoolFormer as an encoder to extract modal feature information using global pooling. VT-UNet \cite{peiris2022vtunet} proposes a Transformer architecture for volume segmentation that maintains a balance in encoding local and global spatial cues. BiTr-Unet \cite{jia2021bitrunet} combines the advantages of CNN and Transformer to achieve accurate segmentation effect by combining models. UNERT \cite{hu2023ertn} Innovatively combines Transformer with U-Net to build feature branches and patch branches and capture complex semantic features and global context information to achieve accurate segmentation. HUT \cite{soh2023hut} consists of two pipelines running in parallel, one based on unet and the other a transformer path that relies on the mid-layer feature mapping of the UNet decoder to improve overall segmentation performance.

\subsection{Multimodal Features Fusion}
Multi-modal medical images of MRI contain richer information than single-modal images. Strategies for different scale feature information based on its multi-encoder and single-decoder structure. CKD-TransBTS\cite{lin2023ckd} Designed parallel structures based on clinical medical knowledge and combined them with interactive attention between associated modes. SMG-BTS \cite{han2023smg-bts} model adopts the structure of three paths, and uses two modes of T1ce and Flair to carry out targeted feature segmentation for different segmentation tasks. Modal features are extracted not only in the network path, but also in the internal modules of the model. Cross-modal transformation (CMFT) and cross-modal fusion (CMFF) constitute a new cross-modal deep learning framework \cite{zhang2021cross}. RFNet \cite{ding2021rfnet} can self-adapt the perception fusion of different mode areas, and effectively aggregate the information of different modes. Zhou et al. \cite{zhou2022tri} proposes three attention fusion modules to strengthen the correlation between modes. F$^{2}$Net \cite{yang2023f2net} proposes a flexible modal fusion framework to realize complementary fusion and feature integration between modes. M2FTrans \cite{shi2023m2ftrans} introduces learnable fusion parameters and masked self-attention to stably establish cross-mode remote dependencies, exploring and fusing cross-mode features via a mode-shielded fusion transformer.

\subsection{Multimodal Features Compression Interaction}
% 既写特征压缩又写特征融合，先写特征压缩，剪枝，+轻量级brats   
With the gradual exploration of BraTS field, the pursuit of lightweight segmentation model with fewer parameters has become a trend. Since SqueezeNet \cite{iandola2016squeezenet} implemented Alexnet \cite{krizhevsky2017alexnet} level accuracy with a smaller CNN architecture, parameters have been reduced by a factor of 50. In the field of BraTS, AST-Net \cite{wang2022astnet} uses lightweight convolution module and shaft space transformer module for feature  extraction in encoder, and MAT \cite{liu2024mat} is a lightweight end-to-end 3D brain tumor segmentation model that  employs an axial attention mechanism to reduce computational requirements and a rectification method to improve  performance on small data sets. ADHDC-Net \cite{liu2023adhdcnet} proposed a lightweight 3D U-Net model that introduced attention mechanism module to address the challenges of difficult recognition of brain tumor regions in medical images, small and irregular targets. MBANet \cite{cao2023mbanet} proposed a 3D lightweight convolutional neural network with 3D multi-branch concerns.

%然后写特征交互
In terms of modal feature interaction, ACMINet \cite{zhuang2022acminet} implements the function of cross-modal feature extraction and interaction to deal with the complex relationship between modes. Xing et.al \cite{xing2024pre} proposes a pre-post interactive learning (PPIL) method that takes the connections of available modes as input to capture more inter-modal correlations. A Post-Interaction component is proposed to sense the importance of all branches and dynamically combine their information. MMCFormer \cite{karimijafarbigloo2024mmcformer} constructs a new missing mode compensation network to solve missing information in an end-to-end manner, incorporating auxiliary tags at the bottleneck stage to simulate the modal interaction between the full and missing mode paths. The difference is that CFCI-Net combines the two modules of modal feature compression and modal interaction, and carries out interactive operation through multi-modal features before and after compression, avoiding the loss of important feature information, and strengthening the integration of multi-modal feature information.

\section{Methodology}
In this section, the overall structure of the model and the various modules of the model are described in detail. The overall model proposed in this paper is shown in Figure 1, the selective mutually exclusive feature fusion (SCFF) module is shown in Figure 2, and the Modal feature channel compression and Interaction (MFCI) transformer is shown in Figure 3.
\begin{figure*}[t!]
\centering{\includegraphics[width=1\linewidth]{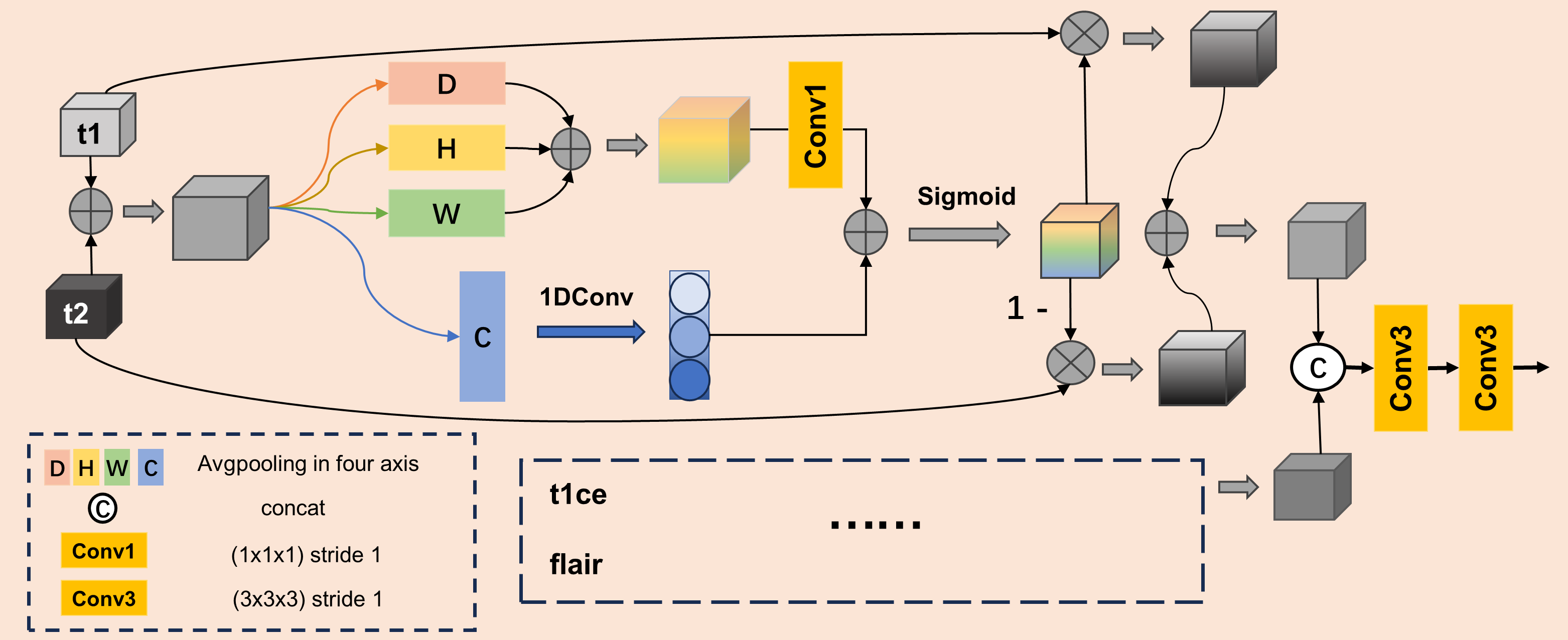}}
\caption{The architecture of the SCFF, it consists of two identical modal feature extraction paths. According to the characteristics of different modes, T1 and T2 are set as a group, and T1ce and Flair are set as a group.}
\label{fig2}
\end{figure*}

\subsection{Selective Complementary Feature Fusion module}
To enhance the fusion of features from distinct modalities, we have conceptualized and implemented the Selective Complementary Feature Fusion (SCFF) module, as depicted in Figure \ref{fig2}. This module is particularly significant in the context of multimodal feature fusion, where the majority of existing models employ straightforward addition or concatenation methods, whichprove to be insufficiently flexible for fusing diverse features, particularly in the realm of MRI multimodal imaging. Features extracted by the encoders from varying modalities share some commonalty yet possess distinct advantages and characteristics. In order to facilitate optimal multimodal feature fusion, we have adopted a strategy based on the principle of complementarity to design the SCFF module. This strategy involves the soft selection processing of feature information from each modality, with a preference for merging features from modalities that exhibit substantial informational discrepancies. This approach ensures that the fusion process incorporates a more comprehensive array of modal features through the employment of complementary weight parameters.

Upon encountering two modal features characterized by disparate feature information, the initial step involves the weight analysis of individual pixel values from each modality. This weight analysis is derived from the interplay of spatial and channel feature information. Subsequently, the spatial weight features along the length, width, and height dimensions are aggregated via a point-by-point convolutional operation, resulting in the construction of a spatial model's weight graph. To facilitate the model's ability to differentiate feature information across varying channel dimensions, we employ one-dimensional convolutions for interactive operations in the channel feature information dimension. Following this, the spatial and channel features are amalgamated, with the corresponding sets of mutually exclusive weight parameters being computed. Finally, the integration of each modality and its corresponding weight parameters is achieved through fusion learning, with the fused feature information being integrated via convolutional operations to accomplish the complementary feature fusion of multi-modes and to optimize the full feature extraction effect. For the fusion of four distinct MRI modalities, we have conducted soft selection and addition operations utilizing T1 and T2, as well as T1ce and Flair. Upon the concatenation of the two sets of modal features, feature splicing is executed, followed by a convolutional neural network (CNN) operation to enhance the feature information of locally significant lesions.
 
For the modal feature vector of X$_{T1/T2/T1ce/Flair}$ $\in$ [C, D, H, W], F is the sum of the two modal features whose feature information differs greatly, the spatial feature weights are obtained by adding the average values of the three dimensions D, W, and H, and then by point-by-point convolution.
\begin{equation}
F^{D} = AVGPOOL(F), F^{D}\in R^{C \times D \times 1 \times 1},
\label{eq1}
\end{equation}
\begin{equation}
F^{H} = AVGPOOL(F), F^{H}\in R^{C \times 1 \times H \times 1},
\label{eq2}
\end{equation}
\begin{equation}
F^{W} = AVGPOOL(F), F^{W}\in R^{C \times 1 \times 1 \times W},
\label{eq3}
\end{equation}

\begin{figure*}[t!]
\centering{\includegraphics[width=1\linewidth]{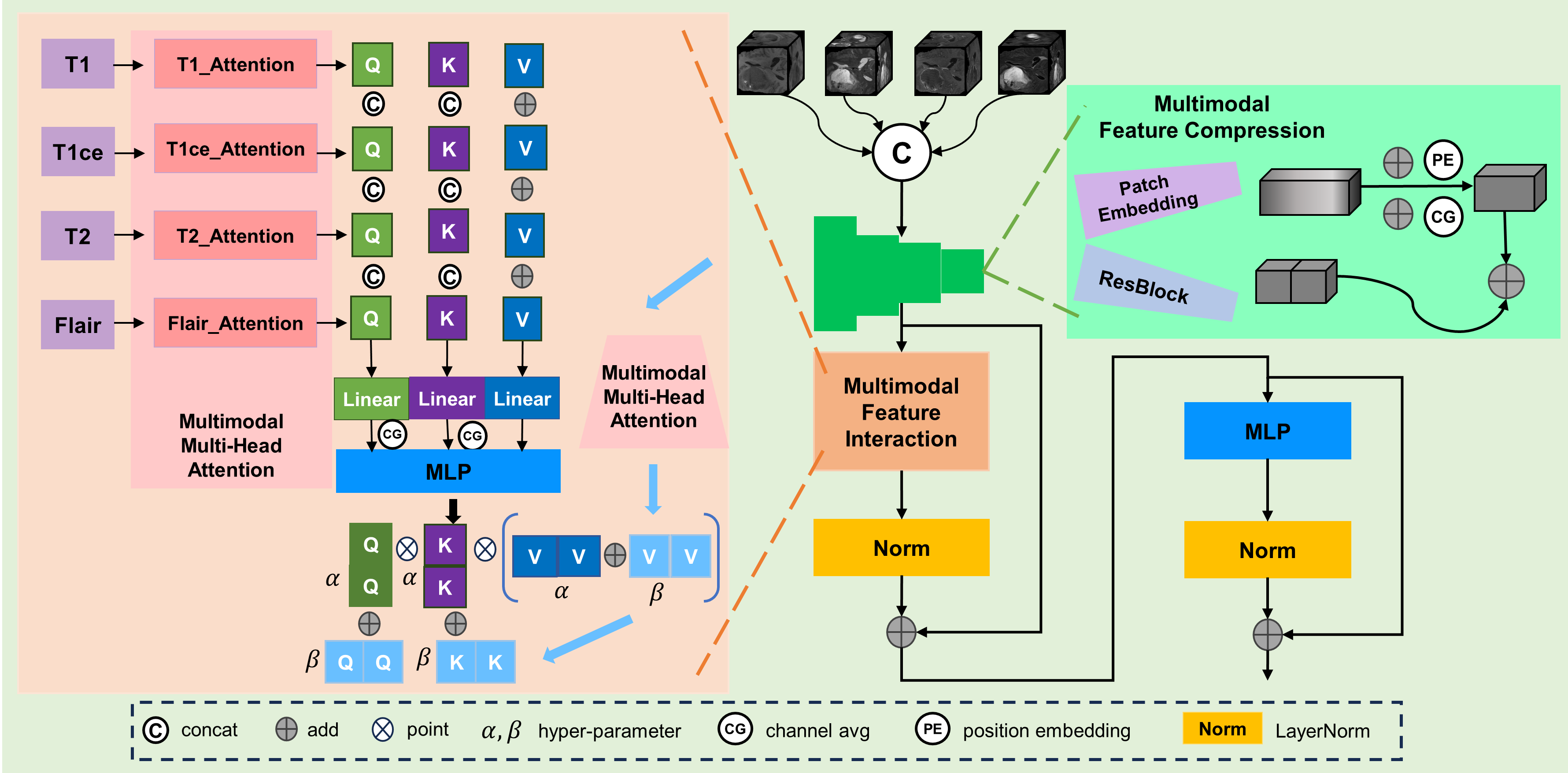}}
\caption{The architecture of the modal feature compression interaction (MFCI) transformer, it consists of a modal feature compression module (MFC) and a modal feature interaction module (MFI).}
\label{fig3}
\end{figure*}
Inspired by ECA-Net \cite{wang2020eca}, our proposed method incorporates an interactive weight convolution operation for channel dimension manipulation. In the local interaction phase, the feature information acquired from diverse modalities by the encoder at varying levels is synergistically refined through interactions with their neighboring features. This facilitates the differentiation of salient lesion characteristics within the feature maps, thereby optimizing the segmentation performance across distinct anatomical regions utilizing the unique attributes of each modality. Channel-wise local attention is achieved via a 1D convolutional operation, establishing a nonlinear mapping relationship between the convolutional kernel parameter K and the channel dimension C, denoted as K = $\varphi$ (C). In the model decoder presented in this study, the channel count is an integer power of 2, resulting in the following relationship:
\begin{equation}
K = \varphi (C) =  \left [ \frac{\log _{2} ^{C}}{\gamma } +  \frac{b}{\gamma } \right ] _{odd} (\gamma =2, b=1),
\label{eq4}
\end{equation}

The operations on the channel dimension are as follows.
\begin{equation}
F^{C} = AVGPOOL(F), F^{C}\in R^{C \times 1 \times 1 \times 1},
\label{eq5}
\end{equation}
\begin{equation}
F^{C} = 1DConv(F^{C}),
\label{eq6}
\end{equation}

The features spatial information and channel characteristics are integrated.
\begin{equation}
F = Sigmoid(F^{Sptial} + F^{C}),
\label{eq7}
\end{equation}
\begin{equation}
Z = Concat((T1 \times F_{T1+T2} + T2 \times (1-F_{T1+T2})), \\
T1ce \times F_{T1ce+Flair} + Flair \times (1-F_{T1ce+Flair})),
\label{eq8}
\end{equation}
\begin{equation}
F^{'} = BN(Conv3(BN(Conv3(Z)))),
\label{eq9}
\end{equation}

BN represent BactchNorm3d, and Conv3 represents the convolution kernel structure of 3x3x3, stride=1.

\subsection{Modal Feature Compression Interactive Transformer}
In the process of extracting feature information from each modality, the encoder integrates the channel features at the lower layers. To this end, we propose the Modal Feature Compression Interaction (MFCI) module. Initially, in response to the issue of dimensionality proliferation in feature maps, we introduce the Modal Feature Compression (MFC) module, which aims to compress features, reduce redundancy, and streamline the model architecture. Subsequently, to guarantee the preservation of critical modal feature information and to enable the model to capture a wealth of multimodal features, we design the Modal Interaction (MFI) module. This module facilitates pre-compression and post-compression feature interactions, as well as inter-modal feature exchanges. Consequently, the model is empowered to engage in robust multimodal learning for MRI image analysis. In the MFI module, we adopt the Transformer architecture, leveraging its superior performance in global attention compared to Convolutional Neural Networks (CNNs) \cite{chen2023weakly}. This enhancement confers a heightened capacity to delineate tumor margins and crucial focal regions, thereby yielding precise tumor segmentation outcomes.

\subsubsection{Modal Feature Compression}
To accomplish efficient modal feature compression, the Modal Feature Compression (MFC) module integrates spatial and sequential feature compression strategies. For spatial feature representation, we employ ResNet blocks to compress features, thereby mitigating the issue of vanishing gradients that can occur at lower levels of the model. Regarding sequential feature information, where each channel's feature data is encoded within its spatial pixel coordinates, it is observed that if the average feature values between two channels are equivalent, the channels may harbor identical or closely related feature subsets. The MFC module leverages this property by utilizing the average feature values of each channel as indexing information, in conjunction with location coding. This approach enables the filtering of redundant and duplicated information within the sequential feature dimensions, enhancing the model's capacity for subsequent processing. Furthermore, the integration of sequential feature information at this stage facilitates the preparation for the Modal Feature Interaction (MFI) module, which operates downstream in the architecture.
\begin{equation}
F = ConCat(T1,T1ce,T2,Flair),
\label{eq10}
\end{equation}
\begin{equation}
F_{sequence} = Patch\_Embedding(F) + Channel_{avg}(F) + PE,
\label{eq11}
\end{equation}
\begin{equation}
F_{c} = Linear(F_{sequence}) + ResBlock(F),
\label{eq12}
\end{equation}

Where, F is the feature information after the concatenation of the four modes T1, T1ce, T2 and Flair, PE is the position coding, F$_{sequence}$ is the sequence feature information, and F$_{c}$ is the mode feature information after compression by the MFC module.
\subsubsection{Modal Feature Interaction}
To enable robust multi-modal feature interaction within the MFCI Transformer and to safeguard against the loss of critical feature information during the compression of spatial and sequential features, we introduce a Modal Feature Interaction (MFI) module subsequent to the Modal Feature Compression (MFC) stage. This module is realized through a sophisticated multi-modal interaction attention mechanism. Here, the MFI is tasked with conducting multi-modal feature interactions on the sequential feature information, and it fuses the feature data extracted by the multi-modal encoder along the sequential axis based on the modal attention weights. This process facilitates comprehensive multi-modal feature interaction both before and after compression, as well as across different levels of the sequential feature hierarchy. Within the interactive multi-head attention mechanism, the query and key matrices are pivotal components that determine the distribution weights of the sequence vector information captured by interactive multi-head attention. The value matrix, on the other hand, is responsible for extracting relevant information from the multi-modal features, particularly focusing on the value vector information of the whole tumor (WT), tumor core (TC), and enhanced tumor (ET) \cite{zhou2020onenet}. Consequently, in the MFI module, we maintain consistent operations for both the query (Q) and key (K) matrices \cite{vaswani2017transformer}. Since different modalities will yield distinct query and key values for the same set of value features, we employ feature concatenation operations for the Q and K matrices of diverse modalities. Given that the segmentation objectives of each modality are equivalent, the WT, TC, and ET tasks must be differentiated, necessitating a uniform value (V) matrix. Feature addition operations are thus adopted in the MFI to facilitate the integration of these diverse modalities.

\begin{equation}
F_{Q} = Concat(Q_{T1}, Q_{T1ce}, Q_{T2}, Q_{Flair}),
\label{eq13}
\end{equation}
\begin{equation}
F_{K} = Concat(K_{T1}, K_{T1ce}, K_{T2}, K_{Flair}),
\label{eq14}
\end{equation}
\begin{equation}
F_{V} = Add(V_{T1}, V_{T1ce}, V_{T2}, V_{Flair}),
\label{eq15}
\end{equation}

In order to make the model integrate the feature information of Q, K and V matrices extracted by various modes, inspired by the attention of multiple co-operators \cite{cordonnier2020collaborate_head}, we raise and reduce dimension of Q, K and V matrices. Specifically, in MFI, we reduce dimension of Query and Key matrices, and increase dimension of Value matrix. Then the matrix operation is carried out on the same dimension level, and further feature learning is carried out through MLP to realize the effective interaction of multi-modal multi-head attention. At the same time, the Query, Key and Value matrix of the feature information after feature compression is extracted from MFCI transformer, and the Query, Key and Value matrix before and after feature compression is combined with hyperparameters $\alpha$ and $\beta$, so as to maintain the important feature information before and after multi-modal feature compression. At the same time, in the hierarchical exchange, The sequence feature information is always dominated by the feature information of the current layer. Combined with the sequence features of the previous layer, this enhances the model's learning and extraction ability for the three segmentation tasks WT, TC and ET, prevents the multiple attention of the model from appearing redundant information again due to excessive attention in a single layer, and prevents the loss of important value information of the segmentation task. Efficient feature compression results are obtained. The specific calculation process is as follows,
\begin{equation}
F_{Q} = MLP(Linear_{Q} (F_{Q} + Channel_{avg} (F_{Q})), \quad
Q = \alpha F_{Q} + \beta Q_{F^{'}},
\label{eq16}
\end{equation}
\begin{equation}
F_{K} = MLP(Linear_{K} (F_{K} + Channel_{avg} (F_{K})), \quad
K = \alpha F_{K} + \beta K_{F^{'}},
\label{eq17}
\end{equation}
\begin{equation}
F_V = MLP(Linear_{V} (F_V)),
\label{eq18}
\end{equation}
\begin{equation}
Attention = Softmax(\frac{Q \times K}{\sqrt{d_{K}}} )\times  (\alpha  F_{V}+\beta  V)
\label{eq19}
\end{equation}

Among them, the Q $_{T1}$, Q$_{T1ce}$,  Q$_{T2}$, Q$_{Flair}$ respectively corresponding to various parts of the four patterns and characteristics of matrix. Linear represents linear layer operation, MLP represents multi-layer perceptron, Channel$_{avg}$ represents the average value of channel features, and V represents the full modal value matrix after modal feature compression.

\section{Experiment}
\subsection{Dataset and Evaluation Metrics}
\subsubsection{BraTS Challenge 2019 and 2020 Dataset}
In this paper, the BraTS2019 and 2020 datasets are used, mainly from the Brain Tumor Segmentation Challenge \cite{menze2014brats}. The data set of BraTS2019 has 335 cases, and the data set of BraTS2020 has 369 cases. The images of each patient include three-dimensional MRI images of T1, T2, T1ce and Flair modes and their corresponding GT mask images, with a resolution of 155×240×240 for each image. The three tasks that need to be segmented are the whole tumor (WT), the tumor core (TC), and the enhanced tumor (ET) region. GT mask contains four categories of labels: 0, 1, 2 and 4, respectively: background, necrotic tumor core, peri-tumor swelling and enhanced tumor. This paper uses the official verification website to benchmark the segmentation model. The official verification data set of BraTS2019 has 125 cases, and the official verification data set of BraTS2020 also has 125 cases.

\begin{table*}[t!]
\centering
\caption{Comparison of Segmentation Performance on \textbf{BRATS2019} Validation Dataset. The Bolded Font Denotes Best Results The Underlined Font Denotes The Second Results}
\renewcommand{\arraystretch}{1} % 增加行高
\setlength{\tabcolsep}{10pt} % 设置列间距
{\footnotesize% 调整字体大小
\begin{tabular}{c|cccc|cccc}
\hline
\multirow{2}{*}{\textbf{Method}} & \multicolumn{4}{c|}{\textbf{Dice Score (\%)}}                                   & \multicolumn{4}{c}{\textbf{Hausdorff Dist. (mm)}}                            \\ \cline{2-9} 
                                 & \textbf{ET} & \textbf{WT} & \multicolumn{1}{c|}{\textbf{TC}}    & \textbf{Avg.} & \textbf{ET} & \textbf{WT} & \multicolumn{1}{c|}{\textbf{TC}} & \textbf{Avg.} \\ \hline
3D U-net \cite{cciccek20163dunet}               & 70.86       & 87.38       & \multicolumn{1}{c|}{72.48}          & 76.91         & 5.062       & 9.432       & \multicolumn{1}{c|}{8.719}       & 7.738         \\
Attention U-Net \cite{oktay1804attentionunet}         & 75.96       & 88.81       & \multicolumn{1}{c|}{77.20}          & 80.66         & 5.202       & 7.756       & \multicolumn{1}{c|}{8.258}       & 7.072         \\
V-Net \cite{milletari2016vnet}                  & 73.89       & 88.73       & \multicolumn{1}{c|}{76.56}          & 79.72         & 6.131       & 6.256       & \multicolumn{1}{c|}{8.705}       & 7.031         \\
CANet \cite{liu2021canet}                  & 75.90       & 88.50       & \multicolumn{1}{c|}{\textbf{85.10}} & 83.17         & 4.809       & 7.091       & \multicolumn{1}{c|}{8.409}       & 6.770         \\
Tunet \cite{vu2020tunet}                  & 78.42       & 90.34       & \multicolumn{1}{c|}{81.12}          & 83.29         & 3.700       & \textbf{4.320}       & \multicolumn{1}{c|}{\underline{6.280}}       & {\underline{4.767}}   \\
AMMGS \cite{liu2023ammgs}              & 76.75       & 89.25       & \multicolumn{1}{c|}{81.10}          & 82.37         & 5.180       & 8.219       & \multicolumn{1}{c|}{7.231} & 6.877         \\
PANet \cite{zhao2022panet}                  & 78.17       & {\underline{90.54}} & \multicolumn{1}{c|}{82.98}          & 83.90         & {\underline{3.453}}       & 4.975       & \multicolumn{1}{c|}{6.852}       & 5.093         \\
TransBTS \cite{wenxuan2021transbts}               & {\underline{78.93}} & 90.00       & \multicolumn{1}{c|}{81.94}          & {\underline{83.62}}         & 3.736       & 5.644       & \multicolumn{1}{c|}{6.049}       & 5.143         \\
SegTran(i3d) \cite{li2105segtran(i3d)}                & 74.00       & 89.50       & \multicolumn{1}{c|}{81.70}          & 81.73         & --           & --           & \multicolumn{1}{c|}{--}           & --             \\
TransUNet \cite{chen2021transunet}              & 78.17       & 89.48       & \multicolumn{1}{c|}{78.91}          & 82.19         & 4.832       & 6.667       & \multicolumn{1}{c|}{7.365}       & 6.288         \\
Swin-UNet \cite{cao2022swinunet}              & 78.49       & 89.38       & \multicolumn{1}{c|}{78.75}          & 82.21         & 6.925       & 7.505       & \multicolumn{1}{c|}{9.260}       & 7.897         \\ \hline
CFCI-Net(Ours)                & \textbf{79.07}        & \textbf{90.61}        & \multicolumn{1}{c|}{\underline{83.38}}               & \textbf{84.35}              & \textbf{3.384}          & {\underline{4.567}}             & \multicolumn{1}{c|}{ \textbf{6.126}}            & \textbf{4.692}              \\ \hline
\end{tabular}
}
\label{table1}
\end{table*}

\begin{table*}[t!]
\centering
\caption{Comparison of Segmentation Performance on \textbf{BRATS2020} Validation Dataset. The Bolded Font Denotes Best Results The Underlined Font Denotes The Second Results}
\renewcommand{\arraystretch}{1} % 增加行高
\setlength{\tabcolsep}{10pt} % 设置列间距
{\footnotesize% 调整字体大小
\makebox[\textwidth][c]{ 
\begin{tabular}{c|cccc|cccc|c}
\hline
\multirow{2}{*}{\textbf{Method}} & \multicolumn{4}{c|}{\textbf{Dice Score (\%)}}                                         & \multicolumn{4}{c|}{\textbf{Hausdorff Dist. (mm)}}                  & \multirow{2}{*}{\textbf{Param. (M)}}                    \\ \cline{2-9} 
                                 & \textbf{ET}    & \textbf{WT}    & \multicolumn{1}{c|}{\textbf{TC}}    & \textbf{Avg.} & \textbf{ET}     & \textbf{WT}    & \multicolumn{1}{c|}{\textbf{TC}}    & \textbf{Avg.} \\ \hline
3D U-Net \cite{cciccek20163dunet}                        & 68.76          & 84.11          & \multicolumn{1}{c|}{79.06}          & 77.31           & 50.983          & 13.366         & \multicolumn{1}{c|}{13.607}         & 25.985      & 16.90      \\
Attention U-Net \cite{oktay1804attentionunet}                        & 71.83          & 85.57          & \multicolumn{1}{c|}{75.96}          & 77.79           & 32.940          & 11.910        & \multicolumn{1}{c|}{19.430}         & 21.437      & 22.76      \\
MDNet \cite{vu2021mdnet}                           & 77.17          & 90.55          & \multicolumn{1}{c|}{82.67}          & 83.46          & 27.040          & 4.990          & \multicolumn{1}{c|}{8.630}          & 13.553          & --    \\
V-Net \cite{milletari2016vnet}                           & 68.97          & 86.11          & \multicolumn{1}{c|}{77.90}          & 77.66              & 43.518          & 14.499         & \multicolumn{1}{c|}{16.153}         & 24.723            & 69.30  \\
nnUNet \cite{isensee2021nnunet}                          & 78.88          & 90.38          & \multicolumn{1}{c|}{82.50}          & 83.92              & 32.740          & 5.350          & \multicolumn{1}{c|}{11.780}         & 16.623          & 28.50     \\
DAUnet \cite{feng2024daunet}                          & 78.60          & 89.80          & \multicolumn{1}{c|}{83.00}          & 83.80              & 27.600          & 5.400          & \multicolumn{1}{c|}{9.800}          & 14.267      & --   \\
ADHDC-Net \cite{liu2023adhdcnet}                       & 78.01          & 89.99          & \multicolumn{1}{c|}{83.31}          & 83.77              & 29.340          & 5.250          & \multicolumn{1}{c|}{9.820}          & 14.803          & 0.3    \\
PANet \cite{zhao2022panet}                           & 78.40          & \textbf{90.90} & \multicolumn{1}{c|}{83.10}          & {\underline{84.13}}              & 26.350          & \textbf{4.020}          & \multicolumn{1}{c|}{\underline{6.510}}          & 12.293           & 19.23   \\
TransBTS \cite{wenxuan2021transbts}                        & 78.73          & 90.09          & \multicolumn{1}{c|}{81.73}          & 83.52              & \textbf{17.947} & 4.964          & \multicolumn{1}{c|}{9.769}          & {\underline{10.893}}        & 32.99      \\
UNETR \cite{hatamizadeh2022unetr}                        & 71.18          & 88.30          & \multicolumn{1}{c|}{75.85}          & 78.44              & 34.460   & 8.180          & \multicolumn{1}{c|}{10.630}          & 17.757        & 102.12      \\
AugTransU-Net \cite{zhang2024augtransunet}                   & 78.60          & 89.80          & \multicolumn{1}{c|}{81.90}          & 83.43              & 24.310          & 5.560          & \multicolumn{1}{c|}{9.560}          & 13.143             & --  \\
SwinBTS \cite{jiang2022swinbts}                         & 77.40          & 89.10          & \multicolumn{1}{c|}{80.30}          & 82.27              & 26.840          & 8.560          & \multicolumn{1}{c|}{15.780}         & 17.060         & 35.70     \\
MixUNet \cite{li2023mixunet}                         & 71.21          & 90.43          & \multicolumn{1}{c|}{80.75}          & 80.80              & 37.865          & 4.792          & \multicolumn{1}{c|}{7.174}         & 16.610          & 2.2    \\
MMTSN \cite{liu2020mmtsn}                           & 76.37          & 88.23          & \multicolumn{1}{c|}{80.12}          & 81.57              & 21.390          & 6.490          & \multicolumn{1}{c|}{6.680}          & 11.520             & -- \\ \hline
CFCI-Net(Ours)                         & \textbf{79.56} & {\underline{90.76}}    & \multicolumn{1}{c|}{\textbf{84.04}} & \textbf{84.79}       & {\underline{20.695}}    & {\underline{4.377}} & \multicolumn{1}{c|}{\textbf{6.252}} & \textbf{10.441}       & 10.61  \\ \hline
\end{tabular}
}}
\label{table2}
\end{table*}
\subsubsection{Evaluation Metrics}
In this paper, Dice score, Sensitivity, Specificity and Hausdorff95 distance were used to evaluate the segmentation accuracy of the model, respectively.
Where TP is the number of samples in which the positive example is predicted as a positive example, TN is the number of samples in which the positive example is predicted as a false example, FP is the number of samples in which the false example is predicted as a positive example, and FN is the number of samples in which the false example is predicted as a false example.
\begin{equation}
Dice =  \frac{2\times TP}{2\times TP+FN+FP} 
\label{eca1}
\end{equation}
\begin{equation}
Sensitivity =  \frac{TP}{TP+FN} 
\label{eca2}
\end{equation}
\begin{equation}
Specificity = \frac{TP}{TP+FP} 
\label{eca3}
\end{equation}
\begin{equation}
D_{Hausdorff}(A, B) = max(h(A, B), h(B, A))
\label{eca4}
\end{equation}

\begin{figure*}[t!]
\centering
%t1
\begin{minipage}{0.07\linewidth}
 	\centerline{\includegraphics[width=\textwidth]{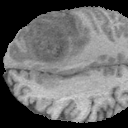}}
 	\vspace{3pt}
 	\centerline{\includegraphics[width=\textwidth]{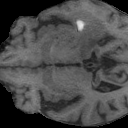}}
 	\vspace{3pt}
 	\centerline{\includegraphics[width=\textwidth]{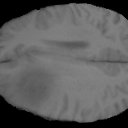}}
 	\vspace{3pt}
 	\centerline{\includegraphics[width=\textwidth]{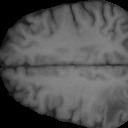}}
 	\vspace{3pt}
 	\centerline{\includegraphics[width=\textwidth]{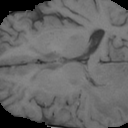}}
 	\vspace{3pt}
 	\centerline{\includegraphics[width=\textwidth]{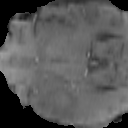}}
 	\vspace{3pt}
  \centerline{\includegraphics[width=\textwidth]{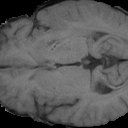}}
 	\vspace{3pt}
  \centerline{\includegraphics[width=\textwidth]{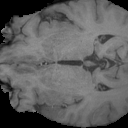}}
 	\vspace{3pt}
  \centerline{\includegraphics[width=\textwidth]{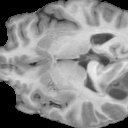}}
 	\vspace{3pt}
  \centerline{\includegraphics[width=\textwidth]{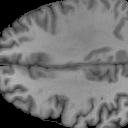}}
 	\centerline{(a)}
 \end{minipage}
 %t2
 \begin{minipage}{0.07\linewidth}
 	\centerline{\includegraphics[width=\textwidth]{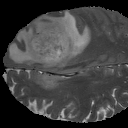}}
 	\vspace{3pt}
 	\centerline{\includegraphics[width=\textwidth]{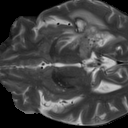}}
 	\vspace{3pt}
 	\centerline{\includegraphics[width=\textwidth]{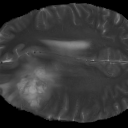}}
 	\vspace{3pt}
 	\centerline{\includegraphics[width=\textwidth]{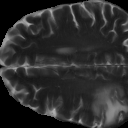}}
 	\vspace{3pt}
 	\centerline{\includegraphics[width=\textwidth]{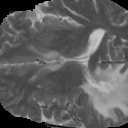}}
 	\vspace{3pt}
 	\centerline{\includegraphics[width=\textwidth]{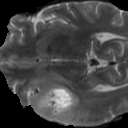}}
 	\vspace{3pt}
  \centerline{\includegraphics[width=\textwidth]{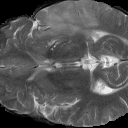}}
 	\vspace{3pt}
  \centerline{\includegraphics[width=\textwidth]{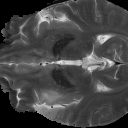}}
 	\vspace{3pt}
  \centerline{\includegraphics[width=\textwidth]{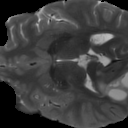}}
 	\vspace{3pt}
  \centerline{\includegraphics[width=\textwidth]{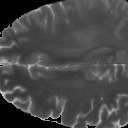}}
 	\centerline{(b)}
 \end{minipage}
 % t1ce
 \begin{minipage}{0.07\linewidth}
 	\centerline{\includegraphics[width=\textwidth]{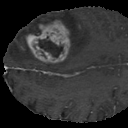}}
 	\vspace{3pt}
 	\centerline{\includegraphics[width=\textwidth]{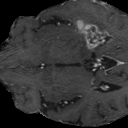}}
 	\vspace{3pt}
 	\centerline{\includegraphics[width=\textwidth]{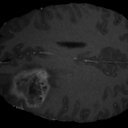}}
 	\vspace{3pt}
 	\centerline{\includegraphics[width=\textwidth]{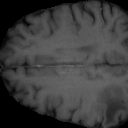}}
 	\vspace{3pt}
 	\centerline{\includegraphics[width=\textwidth]{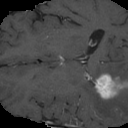}}
 	\vspace{3pt}
 	\centerline{\includegraphics[width=\textwidth]{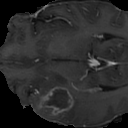}}
 	\vspace{3pt}
  \centerline{\includegraphics[width=\textwidth]{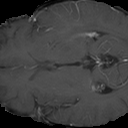}}
 	\vspace{3pt}
  \centerline{\includegraphics[width=\textwidth]{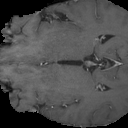}}
 	\vspace{3pt}
  \centerline{\includegraphics[width=\textwidth]{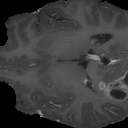}}
 	\vspace{3pt}
  \centerline{\includegraphics[width=\textwidth]{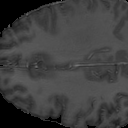}}
 	\centerline{(c)}
 \end{minipage}
 % flair
 \begin{minipage}{0.07\linewidth}
 	\centerline{\includegraphics[width=\textwidth]{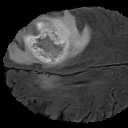}}
 	\vspace{3pt}
 	\centerline{\includegraphics[width=\textwidth]{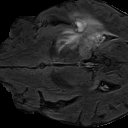}}
 	\vspace{3pt}
 	\centerline{\includegraphics[width=\textwidth]{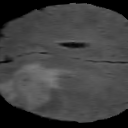}}
 	\vspace{3pt}
 	\centerline{\includegraphics[width=\textwidth]{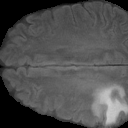}}
 	\vspace{3pt}
 	\centerline{\includegraphics[width=\textwidth]{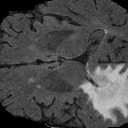}}
 	\vspace{3pt}
 	\centerline{\includegraphics[width=\textwidth]{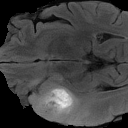}}
 	\vspace{3pt}
  \centerline{\includegraphics[width=\textwidth]{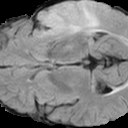}}
 	\vspace{3pt}
  \centerline{\includegraphics[width=\textwidth]{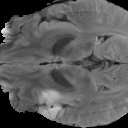}}
 	\vspace{3pt}
  \centerline{\includegraphics[width=\textwidth]{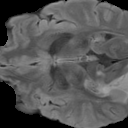}}
 	\vspace{3pt}
  \centerline{\includegraphics[width=\textwidth]{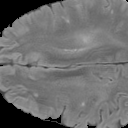}}
 	\centerline{(d)}
 \end{minipage}
 %GT
 \begin{minipage}{0.07\linewidth}
 	\centerline{\includegraphics[width=\textwidth]{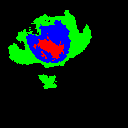}}
 	\vspace{3pt}
 	\centerline{\includegraphics[width=\textwidth]{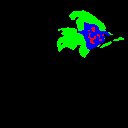}}
 	\vspace{3pt}
 	\centerline{\includegraphics[width=\textwidth]{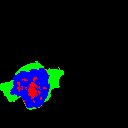}}
 	\vspace{3pt}
 	\centerline{\includegraphics[width=\textwidth]{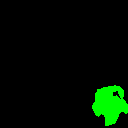}}
 	\vspace{3pt}
 	\centerline{\includegraphics[width=\textwidth]{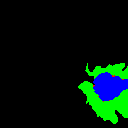}}
 	\vspace{3pt}
 	\centerline{\includegraphics[width=\textwidth]{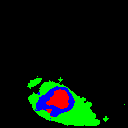}}
 	\vspace{3pt}
  \centerline{\includegraphics[width=\textwidth]{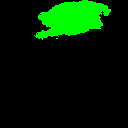}}
 	\vspace{3pt}
  \centerline{\includegraphics[width=\textwidth]{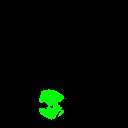}}
 	\vspace{3pt}
  \centerline{\includegraphics[width=\textwidth]{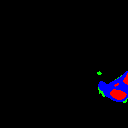}}
 	\vspace{3pt}
  \centerline{\includegraphics[width=\textwidth]{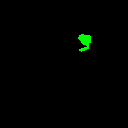}}
 	\centerline{(e)}
 \end{minipage}
 %unet
 \begin{minipage}{0.07\linewidth}
 	\centerline{\includegraphics[width=\textwidth]{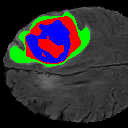}}
 	\vspace{3pt}
 	\centerline{\includegraphics[width=\textwidth]{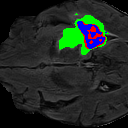}}
 	\vspace{3pt}
 	\centerline{\includegraphics[width=\textwidth]{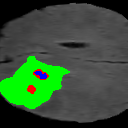}}
 	\vspace{3pt}
 	\centerline{\includegraphics[width=\textwidth]{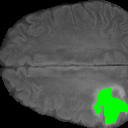}}
 	\vspace{3pt}
 	\centerline{\includegraphics[width=\textwidth]{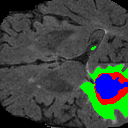}}
 	\vspace{3pt}
 	\centerline{\includegraphics[width=\textwidth]{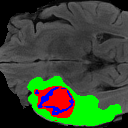}}
 	\vspace{3pt}
  \centerline{\includegraphics[width=\textwidth]{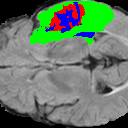}}
 	\vspace{3pt}
  \centerline{\includegraphics[width=\textwidth]{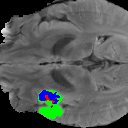}}
 	\vspace{3pt}
  \centerline{\includegraphics[width=\textwidth]{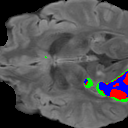}}
 	\vspace{3pt}
  \centerline{\includegraphics[width=\textwidth]{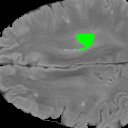}}
 	\centerline{(f)}
 \end{minipage}
 \begin{minipage}{0.07\linewidth}
 	\centerline{\includegraphics[width=\textwidth]{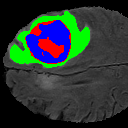}}
 	\vspace{3pt}
 	\centerline{\includegraphics[width=\textwidth]{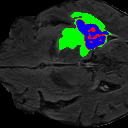}}
 	\vspace{3pt}
 	\centerline{\includegraphics[width=\textwidth]{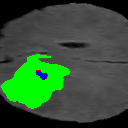}}
 	\vspace{3pt}
 	\centerline{\includegraphics[width=\textwidth]{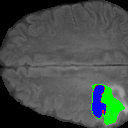}}
 	\vspace{3pt}
 	\centerline{\includegraphics[width=\textwidth]{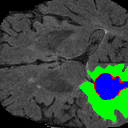}}
 	\vspace{3pt}
 	\centerline{\includegraphics[width=\textwidth]{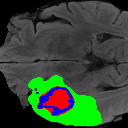}}
 	\vspace{3pt}
  \centerline{\includegraphics[width=\textwidth]{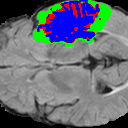}}
 	\vspace{3pt}
  \centerline{\includegraphics[width=\textwidth]{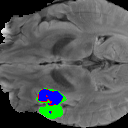}}
 	\vspace{3pt}
  \centerline{\includegraphics[width=\textwidth]{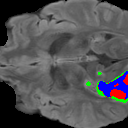}}
 	\vspace{3pt}
  \centerline{\includegraphics[width=\textwidth]{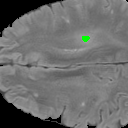}}
 	\centerline{(g)}
 \end{minipage}
 \begin{minipage}{0.07\linewidth}
 	\centerline{\includegraphics[width=\textwidth]{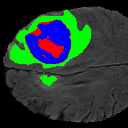}}
 	\vspace{3pt}
 	\centerline{\includegraphics[width=\textwidth]{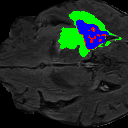}}
 	\vspace{3pt}
 	\centerline{\includegraphics[width=\textwidth]{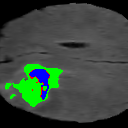}}
 	\vspace{3pt}
 	\centerline{\includegraphics[width=\textwidth]{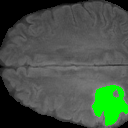}}
 	\vspace{3pt}
 	\centerline{\includegraphics[width=\textwidth]{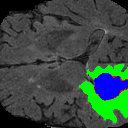}}
 	\vspace{3pt}
 	\centerline{\includegraphics[width=\textwidth]{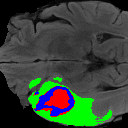}}
 	\vspace{3pt}
  \centerline{\includegraphics[width=\textwidth]{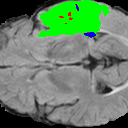}}
 	\vspace{3pt}
  \centerline{\includegraphics[width=\textwidth]{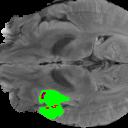}}
 	\vspace{3pt}
  \centerline{\includegraphics[width=\textwidth]{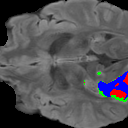}}
 	\vspace{3pt}
  \centerline{\includegraphics[width=\textwidth]{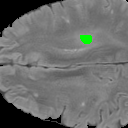}}
 	\centerline{(h)}
 \end{minipage}
 \begin{minipage}{0.07\linewidth}
 	\centerline{\includegraphics[width=\textwidth]{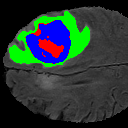}}
 	\vspace{3pt}
 	\centerline{\includegraphics[width=\textwidth]{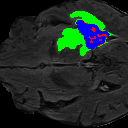}}
 	\vspace{3pt}
 	\centerline{\includegraphics[width=\textwidth]{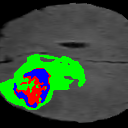}}
 	\vspace{3pt}
 	\centerline{\includegraphics[width=\textwidth]{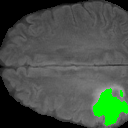}}
 	\vspace{3pt}
 	\centerline{\includegraphics[width=\textwidth]{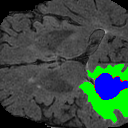}}
 	\vspace{3pt}
 	\centerline{\includegraphics[width=\textwidth]{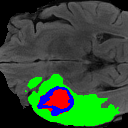}}
 	\vspace{3pt}
  \centerline{\includegraphics[width=\textwidth]{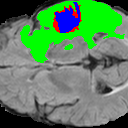}}
 	\vspace{3pt}
  \centerline{\includegraphics[width=\textwidth]{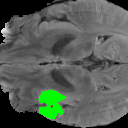}}
 	\vspace{3pt}
  \centerline{\includegraphics[width=\textwidth]{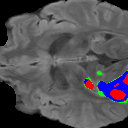}}
 	\vspace{3pt}
  \centerline{\includegraphics[width=\textwidth]{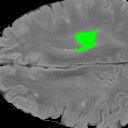}}
 	\centerline{(i)}
 \end{minipage}
 \begin{minipage}{0.07\linewidth}
 	\centerline{\includegraphics[width=\textwidth]{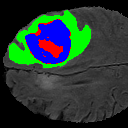}}
 	\vspace{3pt}
 	\centerline{\includegraphics[width=\textwidth]{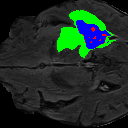}}
 	\vspace{3pt}
 	\centerline{\includegraphics[width=\textwidth]{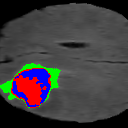}}
 	\vspace{3pt}
 	\centerline{\includegraphics[width=\textwidth]{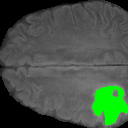}}
 	\vspace{3pt}
 	\centerline{\includegraphics[width=\textwidth]{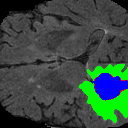}}
 	\vspace{3pt}
 	\centerline{\includegraphics[width=\textwidth]{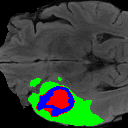}}
 	\vspace{3pt}
  \centerline{\includegraphics[width=\textwidth]{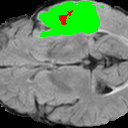}}
 	\vspace{3pt}
  \centerline{\includegraphics[width=\textwidth]{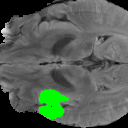}}
 	\vspace{3pt}
  \centerline{\includegraphics[width=\textwidth]{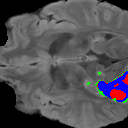}}
 	\vspace{3pt}
  \centerline{\includegraphics[width=\textwidth]{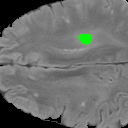}}
 	\centerline{(j)}
 \end{minipage}
 \begin{minipage}{0.07\linewidth}
 	\centerline{\includegraphics[width=\textwidth]{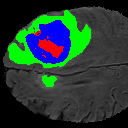}}
 	\vspace{3pt}
 	\centerline{\includegraphics[width=\textwidth]{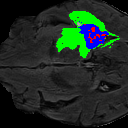}}
 	\vspace{3pt}
 	\centerline{\includegraphics[width=\textwidth]{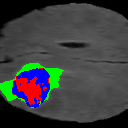}}
 	\vspace{3pt}
 	\centerline{\includegraphics[width=\textwidth]{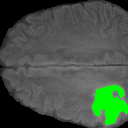}}
 	\vspace{3pt}
 	\centerline{\includegraphics[width=\textwidth]{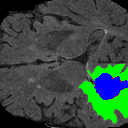}}
 	\vspace{3pt}
 	\centerline{\includegraphics[width=\textwidth]{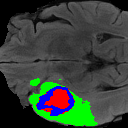}}
 	\vspace{3pt}
  \centerline{\includegraphics[width=\textwidth]{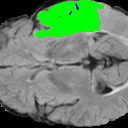}}
 	\vspace{3pt}
  \centerline{\includegraphics[width=\textwidth]{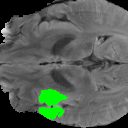}}
 	\vspace{3pt}
  \centerline{\includegraphics[width=\textwidth]{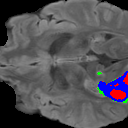}}
 	\vspace{3pt}
  \centerline{\includegraphics[width=\textwidth]{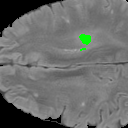}}
 	\centerline{(k)}
 \end{minipage}
 \begin{minipage}{0.07\linewidth}
 	\centerline{\includegraphics[width=\textwidth]{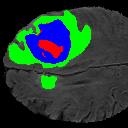}}
 	\vspace{3pt}
 	\centerline{\includegraphics[width=\textwidth]{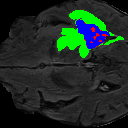}}
 	\vspace{3pt}
 	\centerline{\includegraphics[width=\textwidth]{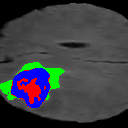}}
 	\vspace{3pt}
 	\centerline{\includegraphics[width=\textwidth]{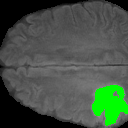}}
 	\vspace{3pt}
 	\centerline{\includegraphics[width=\textwidth]{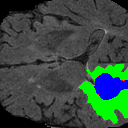}}
 	\vspace{3pt}
 	\centerline{\includegraphics[width=\textwidth]{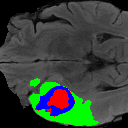}}
 	\vspace{3pt}
  \centerline{\includegraphics[width=\textwidth]{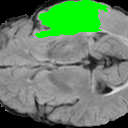}}
 	\vspace{3pt}
  \centerline{\includegraphics[width=\textwidth]{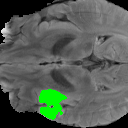}}
 	\vspace{3pt}
  \centerline{\includegraphics[width=\textwidth]{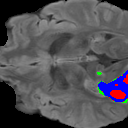}}
 	\vspace{3pt}
  \centerline{\includegraphics[width=\textwidth]{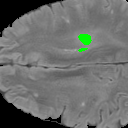}}
 	\centerline{(l)Ours}
 \end{minipage}
\caption{This is a visualization of the comparison experiment on BraTS2020 train dataset, From left to right is (a) T1,  (b) T2,  (c) T1ce,  (d) Flair,  (e) GT,  (f) 3D U-Net,  (g) Attention U-Net,  (h) UNERT,  (i) nnUnet,  (j) TransBTS,  (k) PANet,  (l) CFCI-Net(Ours). To show the segmentation results more clearly, red represents necrotic tumor core (NCR), green represents peritumoral edema (ED), and blue code enhancing tumor (ET).}
\label{fig4}
\end{figure*}
\subsection{Implementation Details}
We implemented all the experiments on PyTorch and MONAI, and trained the model on a workstation using an A40 GPU(48GB). The experiment environment is Ubuntu20.04, Python 3.8, Cuda 11.3. The number of epochs trained by the model is 200, and Dice loss is taken as the objective function. In the training stage, we cut each MRI case into 128x128x128 image blocks for data normalization and standardization. In order to make the data distribution more complex and alleviate the overfitting problem, we used several data enhancement methods, including random scaling, random flipping in three directions, and random cutting. All data enhancement methods are applied to all four modes, with the same Settings. In the testing phase, we set the overlap between patches to 75\% and averaged the predicted results between patches.

\subsection{Compare experiment}
In this section, we will compare BraTS2019 data sets and BraTS2020 various glioma model segmentation of data sets, including some important models. Some of these models are based on convolutional modules, some are based on the Transformer architecture, and some include modal feature fusion strategies and attention mechanisms. Compared with these backbone models and advanced algorithm models, the performance and segmentation ability of CFCI-Net model are better reflected. The experimental results of all models were carried out using the official verification website, and the experimental index evaluation of each model was shown in Table \ref{table1} and Table \ref{table2}. In the comparison experiment, because the BtraTS2020 dataset has more data sets, our visualization experiment is mainly carried out on BraTS2020.

The comparison experiments on the BraTS2019 dataset are shown in Table \ref{table1}. According to the results, CFCI-Net achieved 79.07\%, 90.61\% and 83.38\% effects on the Dice index of ET, WT and TC, and the Hausdorff distance was reduced to 3.384, 4567,6.126. CFCI-Net segmentation model has achieved excellent segmentation results, especially in the evaluation of Dice and average Hausdorff distance. However, the Dice index of CFCI-Net segmentation model in TC is significantly lower than that of CANet. We speculate that this may be due to CANet's possession of complete CC+GC and 5-iteration CGA-CRF achieving competitive performance against the top performer. The Hausdorff distance of TuNet on WT is due to the model proposed by us. We believe that this may be due to the fact that TuNet uses the end-to-end cascading network model architecture, which enables TuNet model to better capture global information and reduce the influence of edge discrete points. On the BraTS2019 dataset, we have also compared with other excellent segmentation models, and the results show that the CFCI-Net model has achieved excellent results, reaching the level of SOTA segmentation.

% 整个模型的消融实验
\begin{table*}[t!]
\caption{About CFCI-Net Ablation Experiment on \textbf{BRATS2020} Validation Dataset. The Bolded Font Denotes Best Results}
\renewcommand{\arraystretch}{1.2} % 增加行高
\setlength{\tabcolsep}{8pt} % 设置列间距
\centering
\makebox[\textwidth][c]{ 
\begin{tabular}{c|ccc|cccc|cccc}
\hline
\multirow{2}{*}{\textbf{Methods}} & \multicolumn{3}{c|}{\textbf{Components}}          & \multicolumn{4}{c|}{\textbf{Dice Score (\%)}}                                          & \multicolumn{4}{c}{\textbf{Hausdorff Dist. (mm)}}                                        \\ \cline{2-12} 
                                  & \textbf{Parallel} & \textbf{MFCI} & \textbf{SMEF} & \textbf{ET}    & \textbf{WT}    & \multicolumn{1}{c|}{\textbf{TC}}    & \textbf{Avg.}  & \textbf{ET}     & \textbf{WT}    & \multicolumn{1}{c|}{\textbf{TC}}    & \textbf{Avg.}   \\ \hline
baseline                          & $\times$                  & $\times$      & $\times$      & 76.74          & 90.01          & \multicolumn{1}{c|}{81.51}          & 82.75          & 29.747          & 5.165          & \multicolumn{1}{c|}{9.892}          & 14.935          \\
baseline(P)                       & $\surd$         & $\times$        & $\times$       & 75.70          & 90.26 & \multicolumn{1}{c|}{82.07}          & 82.67          & 35.604          & 4.943          & \multicolumn{1}{c|}{9.518}          & 16.688          \\
baseline(P)+MFCI                  & $\surd$      & $\surd$       & $\times$      & 77.09          & \textbf{90.82}          & \multicolumn{1}{c|}{83.05}          & 83.65          & 29.834          & 4.379          & \multicolumn{1}{c|}{6.511}          & 13.575          \\
baseline(P)+SCFF                  & $\surd$         & $\times$      & $\surd$       & 77.42          & 90.77          & \multicolumn{1}{c|}{82.72}          & 83.64          & 29.569          & \textbf{4.277} & \multicolumn{1}{c|}{6.610}          & 13.469          \\ \hline
CFCI-Net(Ours)                    & $\surd$         & $\surd$       & $\surd$       & \textbf{79.56} & 90.76          & \multicolumn{1}{c|}{\textbf{84.04}} & \textbf{84.79} & \textbf{20.695} & 4.377          & \multicolumn{1}{c|}{\textbf{6.252}} & \textbf{10.441} \\ \hline
\end{tabular}
}
\label{table3}
\end{table*}

% MFCI的消融实验的图片
\begin{figure*}[t!]
\centering
\subfigure[Flair]
{
\begin{minipage}{0.08\linewidth}
 	\centerline{\includegraphics[width=\textwidth]{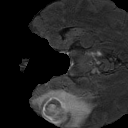}}
 	\vspace{3pt}
 	\centerline{\includegraphics[width=\textwidth]{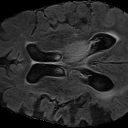}}
 	\vspace{3pt}
 	\centerline{\includegraphics[width=\textwidth]{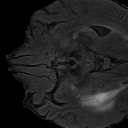}}
 	\vspace{3pt}
 	\centerline{\includegraphics[width=\textwidth]{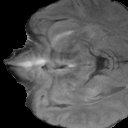}}
 	\vspace{3pt}
 	\centerline{\includegraphics[width=\textwidth]{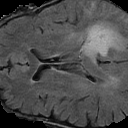}}
 	\vspace{3pt}
 	\centerline{\includegraphics[width=\textwidth]{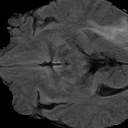}}
 	\vspace{3pt}
  \centerline{\includegraphics[width=\textwidth]{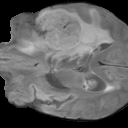}}
 	\vspace{3pt}
  \centerline{\includegraphics[width=\textwidth]{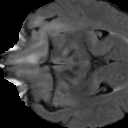}}
  \vspace{3pt}
 \end{minipage}
}
\hspace{-4mm}
\subfigure[GT]{
 \begin{minipage}{0.08\linewidth}
 	\centerline{\includegraphics[width=\textwidth]{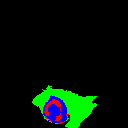}}
 	\vspace{3pt}
 	\centerline{\includegraphics[width=\textwidth]{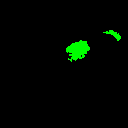}}
 	\vspace{3pt}
 	\centerline{\includegraphics[width=\textwidth]{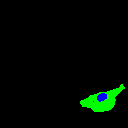}}
 	\vspace{3pt}
 	\centerline{\includegraphics[width=\textwidth]{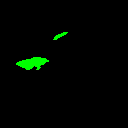}}
 	\vspace{3pt}
 	\centerline{\includegraphics[width=\textwidth]{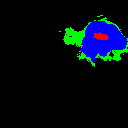}}
 	\vspace{3pt}
 	\centerline{\includegraphics[width=\textwidth]{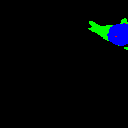}}
 	\vspace{3pt}
  \centerline{\includegraphics[width=\textwidth]{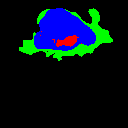}}
 	\vspace{3pt}
  \centerline{\includegraphics[width=\textwidth]{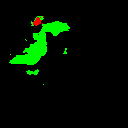}}
  \vspace{3pt}
 \end{minipage}
}
\hspace{-4mm}
\subfigure[baseline]{
 \begin{minipage}{0.08\linewidth}
 	\centerline{\includegraphics[width=\textwidth]{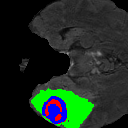}}
 	\vspace{3pt}
 	\centerline{\includegraphics[width=\textwidth]{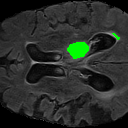}}
 	\vspace{3pt}
 	\centerline{\includegraphics[width=\textwidth]{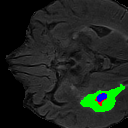}}
 	\vspace{3pt}
 	\centerline{\includegraphics[width=\textwidth]{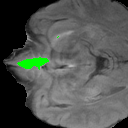}}
 	\vspace{3pt}
 	\centerline{\includegraphics[width=\textwidth]{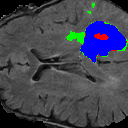}}
 	\vspace{3pt}
 	\centerline{\includegraphics[width=\textwidth]{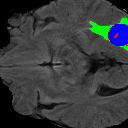}}
 	\vspace{3pt}
  \centerline{\includegraphics[width=\textwidth]{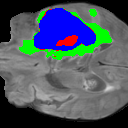}}
 	\vspace{3pt}
  \centerline{\includegraphics[width=\textwidth]{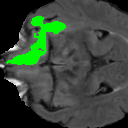}}
  \vspace{3pt}
 \end{minipage}
}
\hspace{-4mm}
\subfigure[baseline\_P]{
\begin{minipage}{0.08\linewidth}
 	\centerline{\includegraphics[width=\textwidth]{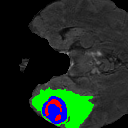}}
 	\vspace{3pt}
 	\centerline{\includegraphics[width=\textwidth]{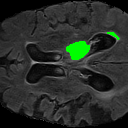}}
 	\vspace{3pt}
 	\centerline{\includegraphics[width=\textwidth]{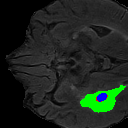}}
 	\vspace{3pt}
 	\centerline{\includegraphics[width=\textwidth]{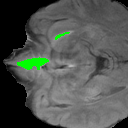}}
 	\vspace{3pt}
 	\centerline{\includegraphics[width=\textwidth]{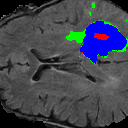}}
 	\vspace{3pt}
 	\centerline{\includegraphics[width=\textwidth]{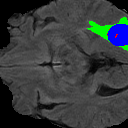}}
 	\vspace{3pt}
  \centerline{\includegraphics[width=\textwidth]{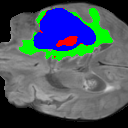}}
 	\vspace{3pt}
  \centerline{\includegraphics[width=\textwidth]{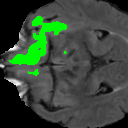}}
  \vspace{3pt}
 \end{minipage}
}
\hspace{-4mm}
\subfigure[MFCI]{
 \begin{minipage}{0.08\linewidth}
 	\centerline{\includegraphics[width=\textwidth]{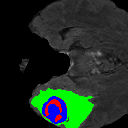}}
 	\vspace{3pt}
 	\centerline{\includegraphics[width=\textwidth]{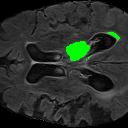}}
 	\vspace{3pt}
 	\centerline{\includegraphics[width=\textwidth]{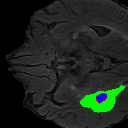}}
 	\vspace{3pt}
 	\centerline{\includegraphics[width=\textwidth]{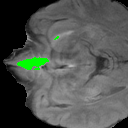}}
 	\vspace{3pt}
 	\centerline{\includegraphics[width=\textwidth]{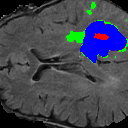}}
 	\vspace{3pt}
 	\centerline{\includegraphics[width=\textwidth]{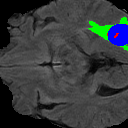}}
 	\vspace{3pt}
  \centerline{\includegraphics[width=\textwidth]{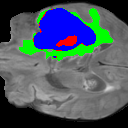}}
 	\vspace{3pt}
  \centerline{\includegraphics[width=\textwidth]{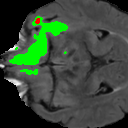}}
  \vspace{3pt}
 \end{minipage}
}
\hspace{-4mm}
\subfigure[SCFF]{
 \begin{minipage}{0.08\linewidth}
 	\centerline{\includegraphics[width=\textwidth]{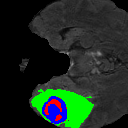}}
 	\vspace{3pt}
 	\centerline{\includegraphics[width=\textwidth]{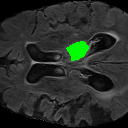}}
 	\vspace{3pt}
 	\centerline{\includegraphics[width=\textwidth]{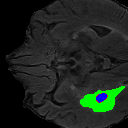}}
 	\vspace{3pt}
 	\centerline{\includegraphics[width=\textwidth]{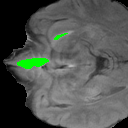}}
 	\vspace{3pt}
 	\centerline{\includegraphics[width=\textwidth]{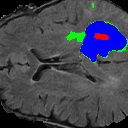}}
 	\vspace{3pt}
 	\centerline{\includegraphics[width=\textwidth]{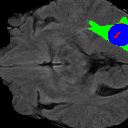}}
 	\vspace{3pt}
  \centerline{\includegraphics[width=\textwidth]{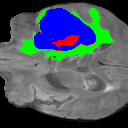}}
 	\vspace{3pt}
  \centerline{\includegraphics[width=\textwidth]{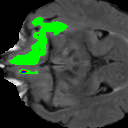}}
  \vspace{3pt}
 \end{minipage}
}
\hspace{-4mm}
\subfigure[CFCI-Net]{
 \begin{minipage}{0.08\linewidth}
 	\centerline{\includegraphics[width=\textwidth]{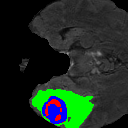}}
 	\vspace{3pt}
 	\centerline{\includegraphics[width=\textwidth]{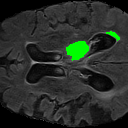}}
 	\vspace{3pt}
 	\centerline{\includegraphics[width=\textwidth]{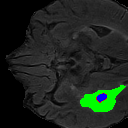}}
 	\vspace{3pt}
 	\centerline{\includegraphics[width=\textwidth]{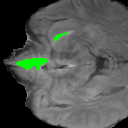}}
 	\vspace{3pt}
 	\centerline{\includegraphics[width=\textwidth]{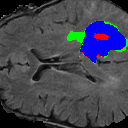}}
 	\vspace{3pt}
 	\centerline{\includegraphics[width=\textwidth]{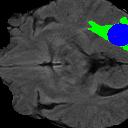}}
 	\vspace{3pt}
  \centerline{\includegraphics[width=\textwidth]{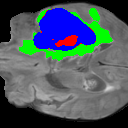}}
 	\vspace{3pt}
  \centerline{\includegraphics[width=\textwidth]{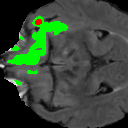}}
  \vspace{3pt}
 \end{minipage}
}
\caption{This is a visualization of the ablation experiment on BraTS2020 train dataset, From left to right is (a)Flair,  (b)GT,  (c)baseline, (d)baseline\_P, (e)MFCI, (f)SCFF, (g)CFCI-Net(Ours). To show the segmentation results more clearly, red represents necrotic tumor core (NCR), green represents peritumoral edema (ED), and blue code enhancing tumor (ET).}
\label{fig5}
\end{figure*}
In the comparison of BraTS2020 validation dataset as shown in Table \ref{table2}, our model EFCI-Net achieved 79.56\%, 90.76\% and 84.04\% results on the Dcie index of the three segmentation tasks of ET, WT and TC, and the Hausdorff distance (mm) was reduced to 20.695, 4.377, 6.252. In addition to being slightly lower than PANet in WT and lower than TransBTS in ET in Hausdorff distance, other indicators are better than other models. In particular, in the mean value of Dice and Hausdorff distance, our model reaches the level of SOTA, leading other segmentation models. This is obviously due to the SCFF module and MFCI transformer proposed by us, which strengthen the extraction and mutually exclusive fusion of different modal features, reduce repetitive and redundant feature information, and achieve accurate segmentation of key focal areas (especially the TC region). As for the excellent performance of PANet on WT, we believe that it may be due to the fact that PANet is the result of cascading network and pays special attention to WT tag, while the Hausdorff distance of TransBTS in ET may be due to its enhanced use of global attention. However, PANet and TransBTS have more parameters than EFCI-Net, resulting in higher complexity of their models. After comparative experiments, remarkable results have been achieved compared with the network based on CNN, Attention U-Net, nnUnet and MDNet, and compared with the model based on Trasnformer, TransBTS, UNETR and SwinBTS. CFCI-Net achieves an advanced level of segmentation. Compared with the lightweight models ADHDC-Net and MixUnet, our model also has excellent results, probably because the lightweight operation makes the model inadequate for the extraction of important features of multi-modes. In addition, although AugTransU-Net focuses on intermodal interaction, its segmentation effect is slightly inferior.

At the same time, in order to display the segmentation results more clearly, we carried out the visualization of the segmentation results. Since the BraTS validation set does not expose the standard GT, we carried out the visualization of the training set. We compared the visualisation results of our CFCI-Net with previous representative methods 3D U-Net, Attention U-Net, UNERT, nnUnet, TransBTS and PANe. In Figure 4, we took the same viewing Angle plane for comparison, and the result was obvious that CFCI-Net showed better segmentation performance in the edge information of tumor areas and important lesion areas. For scattered marginal necrotic tumor core (NCR) features and peritumoral edema (ED) features, it showed good segmentation accuracy. This is attributed to the proposed MFCI transformer for multi-modal feature information compression interactive learning and SCFF module through mutually exclusive weights to effectively extract tumor features of different modes.

% MFCI的消融实验
\begin{table*}[t!]
\makebox[\textwidth][c]{
\begin{threeparttable}
\caption{About MFCI Components Ablation Experiment on \textbf{BRATS2020} Validation Dataset. The Bolded Font Denotes Best Results}
\renewcommand{\arraystretch}{1.2} % 增加行高
\setlength{\tabcolsep}{8pt} % 设置列间距
\centering
\begin{tabular}{c|cc|cccc|cccc|cc}
\hline
\multirow{2}{*}{\textbf{Methods}} & \multicolumn{2}{c|}{\textbf{Components}} & \multicolumn{4}{c|}{\textbf{Dice Score (\%)}}                                          & \multicolumn{4}{c|}{\textbf{Hausdorff Dist. (mm)}}                                       & \multicolumn{2}{c}{\textbf{Complexity}} \\ \cline{2-13} 
                                  & \textbf{MFC}        & \textbf{MFI}       & \textbf{ET}    & \textbf{WT}    & \multicolumn{1}{c|}{\textbf{TC}}    & \textbf{Avg.}  & \textbf{ET}     & \textbf{WT}    & \multicolumn{1}{c|}{\textbf{TC}}    & \textbf{Avg.}   & \textbf{Param.(M)}   & \textbf{GFlops}  \\ \hline
baseline                          & $\times$             & $\times$           & 77.42               & 90.77               & \multicolumn{1}{c|}{82.72}               & 83.64               & 29.569                & 4.227               & \multicolumn{1}{c|}{6.610}               & 13.469                & 5.19                     & 166.04             \\
baseline+MFC                      & $\surd$            & $\times$         & 77.52               & \textbf{90.78}               & \multicolumn{1}{c|}{82.76}               & 83.69               & 26.575                & \textbf{4.122}               & \multicolumn{1}{c|}{6.745}               & 12.481                & 8.21              & 178.42            \\
baseline+MFI                      & $\times$              & $\surd$            & 78.41               & 90.65               & \multicolumn{1}{c|}{82.23}               & 83.76               & 23.704                & 4.174               & \multicolumn{1}{c|}{6.716}               & 11.531                & 12.63                     & 204.55                 \\ \hline
EFCI-Net(Ours)                    & $\surd$              & $\surd$           & \textbf{79.56} & 90.76 & \multicolumn{1}{c|}{\textbf{84.04}} & \textbf{84.79} & \textbf{20.695} & 4.377 & \multicolumn{1}{c|}{\textbf{6.252}} & \textbf{10.441} & 10.61                 & 196.29          \\ \hline
\end{tabular}
\begin{tablenotes}
\item The baseline model here is the result of removing the MFCI module from CFCI-Net.
\end{tablenotes}
\label{table4}
\end{threeparttable}
}
\end{table*}

% SCFF消融实验
\begin{table}[t!]
\centering
\caption{About SCFF Ablation Experiment on \textbf{BRATS2020} Validation Dataset. The Bolded Font Denotes Best Results}
% \tiny\scriptsize
\renewcommand{\arraystretch}{1.1}
\setlength{\tabcolsep}{8pt}
\makebox[\textwidth][c]{
\begin{tabular}{c|clll|clll}
\hline
\multirow{2}{*}{\textbf{Mode Fusion}}                             & \multicolumn{4}{c|}{\textbf{Dice Score (\%)}}                                                                                      & \multicolumn{4}{c}{\textbf{Hausdorff Dist. (mm)}}                                                                                         \\ \cline{2-9} 
                                                                  & \textbf{ET}          & \multicolumn{1}{c}{\textbf{WT}} & \multicolumn{1}{c|}{\textbf{TC}}    & \multicolumn{1}{c|}{\textbf{Avg.}}  & \textbf{ET}                & \multicolumn{1}{c}{\textbf{WT}}    & \multicolumn{1}{c|}{\textbf{TC}}    & \multicolumn{1}{c}{\textbf{Avg.}} \\ \hline
\begin{tabular}[c]{@{}c@{}}T1+T1ce,\\ T2+Flair\end{tabular}       & 78.03                & 90.75                           & \multicolumn{1}{l|}{83.05}          & 83.94                               & \multicolumn{1}{l}{26.626} & 4.670                              & \multicolumn{1}{l|}{6.628}          & 12.641                            \\ \cline{1-1}
\begin{tabular}[c]{@{}c@{}}T1+Flair,\\ T1ce+T2\end{tabular}       & \multicolumn{1}{l}{77.18} & 90.70                            & \multicolumn{1}{l|}{83.02}               & 83.63                                    & \multicolumn{1}{l}{29.845}       & 4.826                                   & \multicolumn{1}{l|}{6.599}               & 13.756                                  \\ \cline{1-1}
\begin{tabular}[c]{@{}c@{}}T1+T2,\\ T1ce+Flair(Ours)\end{tabular} & \textbf{79.56}       & \textbf{90.76}                  & \multicolumn{1}{c|}{\textbf{84.04}} & \multicolumn{1}{c|}{\textbf{84.79}} & \textbf{20.695}            & \multicolumn{1}{c}{\textbf{4.377}} & \multicolumn{1}{c|}{\textbf{6.252}} & \textbf{10.441}                   \\ \hline
\end{tabular}
}
\label{table5}
\end{table}
\subsection{Ablation Studies}
In order to further confirm the effectiveness of our proposed modules, we set corresponding ablation experiments between the overall model and each module. At the same time, in order to prove the rationality of our module design, we designed different ablation experiments for different modules. Because BraTS2020 contains more datasets, ablation experiments are mainly conducted on BraTS2020 datasets. In order to conduct more comprehensive and authoritative analysis, all experimental results are verified by the official website.

\subsubsection{Ablation studies for CFCI-Net}
Here, we will explore the effects and effectiveness of SCFF and MFCI modules on CFCI-Net, and four sets of detailed experiments will be recorded and compared. In order to more effectively evaluate the effects of each CFCI-Net module, we conducted all the ablation experiments using the official validation dataset of BraTS2020 and submitted all the results to the online website for evaluation. In order to more accurately judge the effects of SCFF and MFCI transformer and exclude the gain and influence brought by parallel networks, we used two baseline models at the beginning of the ablation experiment. baseline is a network structure similar to 3D UNet. baseline (P) is a parallel 3D UNet network architecture. The results show that compared with the single network structure, the parallel network achieves better effect on WT and TC, but the effect on ET is reduced. This may be due to the accumulation of a large amount of modal feature information when the parallel network is splicing the end channel of the encoder, which leads to the masking of the features of key focal areas. So it is very important to solve the problem of the proliferation of channel features. The results of the ablation experiment show that MFCI has different degrees of improvement in the three segmentation tasks of ET, WT, and TC. This is because the MFC module in MFCI reduces the redundant and repetitive features of the model, and then combines with the MFI feature interaction module for feature interaction to enhance the fusion learning effect of the model for multi-modal features. SCFF has achieved good results on WT and TC, indicating that the selective mutually exclusive feature fusion module has achieved excellent results on modal feature fusion. Soft selection of mutually exclusive weight parameters is used to realize the complementary effect between different modal features. The CFCI-Net model achieved 79.56\%, 9076\%, 84.04\% Dice and 20.695, 4.377, 6.252 Hausdorff distance on the three segmentation tasks of ET, WT and TC, and achieved excellent results.

\begin{figure}[t!]
\subfigure[Flair]
{
 \begin{minipage}{0.16\linewidth}
 	\centerline{\includegraphics[width=\textwidth]{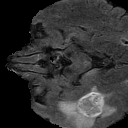}}
 	\vspace{2pt}
 	\centerline{\includegraphics[width=\textwidth]{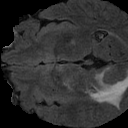}}
 	\vspace{2pt}
 	\centerline{\includegraphics[width=\textwidth]{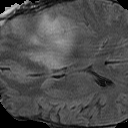}}
 	\vspace{2pt}
 	\centerline{\includegraphics[width=\textwidth]{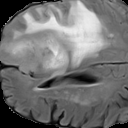}}
 	\vspace{2pt}
  \centerline{\includegraphics[width=\textwidth]{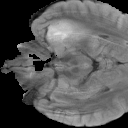}}
 	\vspace{2pt}
  \centerline{\includegraphics[width=\textwidth]{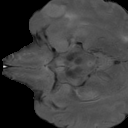}}
 	\vspace{2pt}
  \centerline{\includegraphics[width=\textwidth]{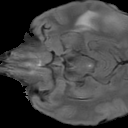}}
 	\vspace{3pt}
 \end{minipage}
}
\hspace{-4mm}
\subfigure[GT]
{
 \begin{minipage}{0.16\linewidth}
 	\centerline{\includegraphics[width=\textwidth]{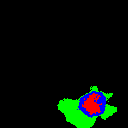}}
 	\vspace{2pt}
 	\centerline{\includegraphics[width=\textwidth]{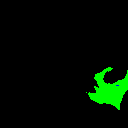}}
 	\vspace{2pt}
 	\centerline{\includegraphics[width=\textwidth]{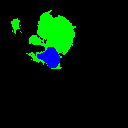}}
 	\vspace{2pt}
 	\centerline{\includegraphics[width=\textwidth]{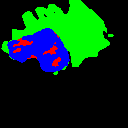}}
 	\vspace{2pt}
  \centerline{\includegraphics[width=\textwidth]{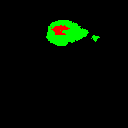}}
 	\vspace{2pt}
  \centerline{\includegraphics[width=\textwidth]{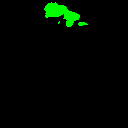}}
 	\vspace{2pt}
  \centerline{\includegraphics[width=\textwidth]{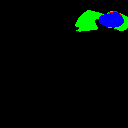}}
 	\vspace{3pt}
 \end{minipage}
}
\hspace{-4mm}
\subfigure[baseline]
{
 \begin{minipage}{0.16\linewidth}
 	\centerline{\includegraphics[width=\textwidth]{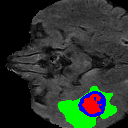}}
 	\vspace{2pt}
 	\centerline{\includegraphics[width=\textwidth]{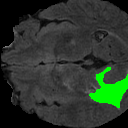}}
 	\vspace{2pt}
 	\centerline{\includegraphics[width=\textwidth]{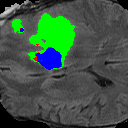}}
 	\vspace{2pt}
 	\centerline{\includegraphics[width=\textwidth]{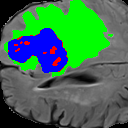}}
 	\vspace{2pt}
  \centerline{\includegraphics[width=\textwidth]{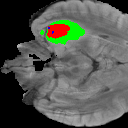}}
 	\vspace{2pt}
  \centerline{\includegraphics[width=\textwidth]{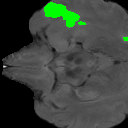}}
 	\vspace{2pt}
  \centerline{\includegraphics[width=\textwidth]{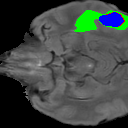}}
 	\vspace{3pt}
 \end{minipage}
}
\hspace{-4mm}
\subfigure[MFC]
{
 \begin{minipage}{0.16\linewidth}
 	\centerline{\includegraphics[width=\textwidth]{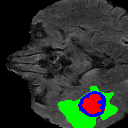}}
 	\vspace{2pt}
 	\centerline{\includegraphics[width=\textwidth]{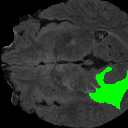}}
 	\vspace{2pt}
 	\centerline{\includegraphics[width=\textwidth]{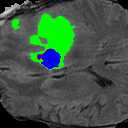}}
 	\vspace{2pt}
 	\centerline{\includegraphics[width=\textwidth]{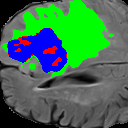}}
 	\vspace{2pt}
  \centerline{\includegraphics[width=\textwidth]{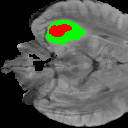}}
 	\vspace{2pt}
  \centerline{\includegraphics[width=\textwidth]{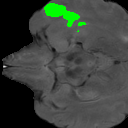}}
 	\vspace{2pt}
  \centerline{\includegraphics[width=\textwidth]{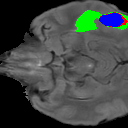}}
 	\vspace{3pt}
 \end{minipage}
}
\hspace{-4mm}
\subfigure[MFI]
{
 \begin{minipage}{0.16\linewidth}
 	\centerline{\includegraphics[width=\textwidth]{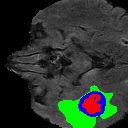}}
 	\vspace{2pt}
 	\centerline{\includegraphics[width=\textwidth]{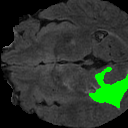}}
 	\vspace{2pt}
 	\centerline{\includegraphics[width=\textwidth]{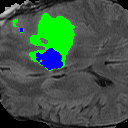}}
 	\vspace{2pt}
 	\centerline{\includegraphics[width=\textwidth]{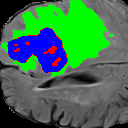}}
 	\vspace{2pt}
  \centerline{\includegraphics[width=\textwidth]{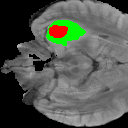}}
 	\vspace{2pt}
  \centerline{\includegraphics[width=\textwidth]{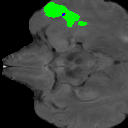}}
 	\vspace{2pt}
  \centerline{\includegraphics[width=\textwidth]{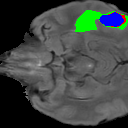}}
 	\vspace{3pt}
 \end{minipage}
}
\hspace{-4mm}
\subfigure[CFCI-Net]
{
 \begin{minipage}{0.16\linewidth}
 	\centerline{\includegraphics[width=\textwidth]{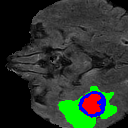}}
 	\vspace{2pt}
 	\centerline{\includegraphics[width=\textwidth]{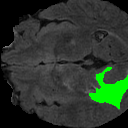}}
 	\vspace{2pt}
 	\centerline{\includegraphics[width=\textwidth]{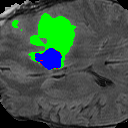}}
 	\vspace{2pt}
 	\centerline{\includegraphics[width=\textwidth]{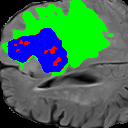}}
 	\vspace{2pt}
  \centerline{\includegraphics[width=\textwidth]{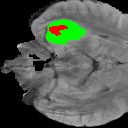}}
 	\vspace{2pt}
  \centerline{\includegraphics[width=\textwidth]{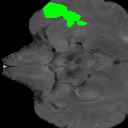}}
 	\vspace{2pt}
  \centerline{\includegraphics[width=\textwidth]{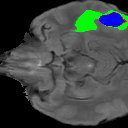}}
 	\vspace{3pt}
 \end{minipage}
}
\caption{This is a visualization of the ablation experiment on BraTS2020 train dataset, From left to right is (a)Flair,  (b)GT,  (c)baseline, (d)MFC, (e)MFI, (f)CFCI-Net(Ours). To show the segmentation results more clearly, red represents necrotic tumor core (NCR), green represents peritumoral edema (ED), and blue code enhancing tumor (ET).}
\label{fig6}
\end{figure}

\begin{table*}[t!]
\centering
\makebox[\textwidth][c]{
\begin{threeparttable}
\caption{About MFCI Layers Ablation Experiment on \textbf{BRATS2020} Validation Dataset. The Bolded Font Denotes Best Results}
% \tiny\scriptsize
\renewcommand{\arraystretch}{1.3}
\setlength{\tabcolsep}{9pt}
\begin{tabular}{c|c|c|cccc|cccc|cc}
\hline
\multirow{2}{*}{\textbf{Stage}}              & \multirow{2}{*}{\textbf{L1}} & \multirow{2}{*}{\textbf{L2}} & \multicolumn{4}{c|}{\textbf{Dice Score (\%)}}                                          & \multicolumn{4}{c|}{\textbf{Hausdorff Dist. (mm)}}                                       & \multicolumn{2}{c}{\textbf{Complexity}} \\ \cline{4-13} 
                                             &                              &                              & \textbf{ET}    & \textbf{WT}    & \multicolumn{1}{c|}{\textbf{TC}}    & \textbf{Avg.}  & \textbf{ET}     & \textbf{WT}    & \multicolumn{1}{c|}{\textbf{TC}}    & \textbf{Avg.}   & \textbf{Param.(M)}   & \textbf{GFlops}  \\ \hline
\multirow{2}{*}{\textbf{L1 \textgreater L2}} & 6                            & 2                            & 77.59          & 90.78          & \multicolumn{1}{c|}{82.56}          & 83.64          & 29.466          & 4.138          & \multicolumn{1}{c|}{6.326}          & 13.310          & 10.31                & 197.50           \\
                                             & 6                            & 4                            & 77.04          & 90.70          & \multicolumn{1}{c|}{83.10}          & 83.61          & 29.636          & 4.439          & \multicolumn{1}{c|}{6.745}          & 13.607          & 10.90                & 200.73           \\
\multirow{2}{*}{\textbf{L1 \textless L2}}    & 4                            & 12                           & 77.27          & \textbf{90.87} & \multicolumn{1}{c|}{83.23}          & 83.79          & 29.610          & \textbf{4.015} & \multicolumn{1}{c|}{9.453}          & 14.359          & 12.98                & 209.22           \\
                                             & 4                            & 16                           & 78.08          & 90.82          & \multicolumn{1}{c|}{83.35}          & 84.08          & 26.668          & 4.110          & \multicolumn{1}{c|}{9.359}          & 13.379          & 14.17                & 215.69           \\
\multirow{3}{*}{\textbf{L1 = L2}}            & 2                            & 2                            & 78.51          & 90.66          & \multicolumn{1}{c|}{82.99}          & 84.05          & 26.546          & 4.766          & \multicolumn{1}{c|}{6.859}          & 12.724          & 9.71                 & 188.61           \\
                                             & 4                            & 4                            & \textbf{79.56} & 90.76          & \multicolumn{1}{c|}{\textbf{84.04}} & \textbf{84.79} & \textbf{20.695} & 4.377          & \multicolumn{1}{c|}{\textbf{6.252}} & \textbf{10.441} & 10.61                & 196.29           \\
                                             & 6                            & 6                            & 77.91          & 90.70          & \multicolumn{1}{c|}{83.20}          & 83.94          & 27.067          & 4.510          & \multicolumn{1}{c|}{9.514}          & 13.697          & 11.50                & 203.96           \\ \hline
\end{tabular}
\begin{tablenotes}
\item L1 is the number of layers of single-mode feature extraction and L2 is the number of layers of multi-mode feature interaction(MFI)
\end{tablenotes}
\label{table6}
\end{threeparttable}
}
\end{table*}
In order to show the segmentation results of the ablation experiment more clearly, we visualized the training set. We visually compared our CFCI-Net results with baseline, baseline\_P, baseline\_P+MFCI, and baseline\_P+SCFF. As shown in FIG. 5, we compared EFCI-Net with MFCI module and SMEF module at the same viewing plane, and the results showed that CFCI-NET did show good segmentation accuracy with the addition of the MFCI module and SCFF module, and performed well in distinguishing normal brain tissue areas from focal areas and in the segmentation of scattered tumor features.

\subsubsection{Ablation studies for SCFF}
In order to verify the rationality of selective mutually exclusive feature fusion, we set up three sets of ablation experiments on different modal fusion strategies to prove that T1 and T2, T1ce and Flair are the most successful fusion strategies. Through the experimental results, we confirmed that the fusion of T1 and T2, and the fusion of T1ce and Flair are the best modal fusion strategies. Since the modal difference of each 
group is maximized in T1 and T2, T1ce and Flair, which can give full play to the role of mutually exclusive selection weights, T1 contains more brain structure information, and more complete brain glioma characteristics can be obtained by combining the edema characteristics and abnormal brain tissue information of T2. T1ce can clearly display the swollen areas of brain tumors and combine with the internal characteristics of tumors shown by Flair to obtain relatively complete brain glioma characteristics []. Therefore, this mutually exclusive feature fusion strategy achieves excellent results. Compared with the previous combination that overemphasizes single advantages (T1 and T1ce, T2 and Flair), it can achieve more accurate segmentation, because it is easy to ignore the advantages of each mode by over-combining the advantages of different modes, thus making the model pay too much attention to single feature information. In this ablation experiment, the fusion strategies of T1 and T1ce, T2 and Flair showed deficiencies in the three segmentation tasks. Therefore, mutually exclusive feature fusion is suitable for MRI multi-modal fusion strategy, and can improve the segmentation effect of key lesion areas.

\subsubsection{Ablation studies for MFCI}
First, as shown in Table \ref{table4}, we conducted the ablation experiment inside the MFCI transformer, and discussed in detail the roles of MFC and MFI. According to the experimental results, the addition of MFC and MFI modules respectively increased the segmentation indexes, indicating that when MFC compressed the feature channels, redundant and repetitive information was reduced, and the influence of outliers of WT on the segmentation results was reduced. The reduction of some channel features did not have adverse effects on the segmentation results of brain glioma. MFI can realize multi-modal feature interactive operation, especially for the segmentation task of important and critical focus such as ET. However, no matter MFC or MFI, the function of the two alone is not as good as the synergistic effect. Using MFC alone without MFI will lead to the omission and loss of key focal region features, and the model complexity will increase. The use of MFI alone without MFC will lead to a large amount of redundant and repetitive information, which will affect the accuracy of model segmentation. Therefore, when MFC and MFI cooperate, the model segmentation effect is the best, and the complexity of the model can be reduced. 
Meanwhile, as shown in Figure \ref{fig6}, through the visualization of the BraTS2020 training data set, it can be seen that the modal feature interaction (MFI) module plays an important role in the localization and recognition of important lesion areas. Modal feature compression (MFC) performs well in processing redundant features and identifying the whole tumor, but the two need to work together. EFCI-Net model can better segment the edge information of brain glioma and the entire tumor region, which helps to improve the segmentation accuracy of the model.

Then, because MFCI achieves accurate segmentation of brain glioma through global capture of multi-modal sequence features, MFCI's design largely follows Transformer's approach to layer processing. Here, we will explore the relationship and proportion between single-modal extraction layers and multi-modal interaction layers. Expect to find the best level design for MFCI. In the ablation experiment shown in Table \ref{table6}, we mainly discussed three situations: Mode\_layer greater than MFI\_layer, Mode\_layer equal to MFI\_layer, and Mode\_layer less than MFI\_layer. According to the experimental results, the best effect is achieved when Mode\_layer is equal to MFI\_layer is equal to Table \ref{table6}. The MFCI transformer is set up for better interactive learning of multi-modal features while reducing redundant parameters. The experimental results show that this module does not perform well on the unbalanced Layers setting, because the unbalanced and unequal Layers setting may cause the model to control the global attention too scattered. Although it shows the Dice index of 90.87\% on WT, it is in the important lesion area ET. The segmentation effect of TC is lower than the parameter setting of Layers balance in L1 and L2, which also indicates that in MFCI, single-mode feature extraction and multi-mode feature interaction are uniformly and cooperatively carried out. For the optimal Layers setting, we found that L1 equals L2 equals 4, the best effect, because in MFCI, the interaction and learning of multi-modal features are processed through sequence features, too little or too much Layers Settings will lead to different degrees of modal features loss and redundancy.

\section{Conclusion}
In this paper, we propose the selective exclusive mode feature fusion and mode feature compression interactive network CFCI-Net. The mode feature compression interaction (MFCI) transformer can deal with the problem of the proliferation of dimensional features in the end channel of the encoder. At the same time, multi-mode learning and feature extraction are carried out through modal interaction. The selective complementary feature Fusion (SCFF) module is the soft processing of two modal features whose feature information is quite different by mutually exclusive weight parameters. We have carried out a large number of verification experiments on BraTS2019 and BraTS2020 to prove the excellent performance of the model. At the same time, a large number of ablation experiments were conducted on the SCFF and MFCI modules to prove the rationality and effectiveness of the module design. Compared with other brain glioma segmentation models, our model achieved advanced segmentation accuracy indicators.

%Bibliography
\section*{References} 
\bibliographystyle{unsrt}  
% \bibliography{references}  
\bibliography{ref}

\end{document}